\documentclass[twocolumn, trackchanges]{aastex6}

\usepackage{graphicx}
\usepackage{amsmath, amssymb}
\usepackage{multirow}
\usepackage[version=3]{mhchem}
\usepackage[normalem]{ulem}
\usepackage{natbib, xfrac}
\bibpunct{(}{)}{;}{a}{}{,}

\graphicspath{{./figures/}}

\newcommand{\reffig}[1]{Fig.~\ref{#1}}
\newcommand{\reftab}[1]{Table~\ref{#1}}

\newcommand{\TWH}{TW~Hya}
\newcommand{\DMT}{DM~Tau}
\newcommand{\ctwoh}{C$_2$H}

\begin{document}

\title{Hydrocarbon emission rings in protoplanetary disks induced by dust evolution}

 \author{Edwin A. Bergin\altaffilmark{1}, Fujun Du\altaffilmark{1}, L. Ilsedore Cleeves\altaffilmark{2},
 G.A. Blake\altaffilmark{3}, K. Schwarz\altaffilmark{1}, R. Visser\altaffilmark{4}, and K. Zhang\altaffilmark{1}}

\altaffiltext{1}{Department of Astronomy,
       University of Michigan,
       311 West Hall, 1085 S. University Ave,
       Ann Arbor, MI 48109, USA}
\altaffiltext{2}{Harvard-Smithsonian Center for Astrophysics, 60 Garden Street, Cambridge, MA 02138}
\altaffiltext{3}{Division of Geological \& Planetary Sciences, MC 150-21,
California Institute of Technology, 1200 E California Blvd,
Pasadena, CA 91125}
\altaffiltext{4}{European Southern Observatory, Karl-Schwarzschild-Str. 2, D-85748, Garching, Germany}



\begin{abstract}%
We report observations of resolved C$_2$H emission rings within the gas-rich protoplanetary disks of TW~Hya and DM~Tau using the Atacama Large Millimeter Array (ALMA).  In each case the emission ring is found to arise at the edge of the observable disk of mm-sized grains (pebbles)  traced by (sub)mm-wave continuum emission.  In addition, we detect a C$_3$H$_2$ emission ring with an identical spatial distribution to C$_2$H in the TW~Hya disk.  This suggests that these are hydrocarbon rings (i.e. not limited to C$_2$H).  Using a detailed thermo-chemical model we show that reproducing the emission from C$_2$H requires a strong UV field and C/O $> 1$ in the upper disk atmosphere and outer disk, beyond the edge of the pebble disk. This naturally arises in a disk where the ice-coated dust mass is spatially stratified due to the combined effects of coagulation, gravitational settling and drift. This stratification causes the disk surface and outer disk to have a greater permeability to UV photons.  
Furthermore the concentration of ices that transport key volatile carriers of oxygen and carbon in the midplane, along with photochemical erosion of CO, leads to an elemental C/O ratio that exceeds unity in the UV-dominated disk.  Thus the motions of the grains, and not the gas, lead to a rich hydrocarbon chemistry in disk surface layers and in the outer disk midplane.  
\end{abstract}



\section{Introduction}

The birth of planetary systems begins with the gravitational collapse of a centrally concentrated core in 
molecule-dominated gas that forms a star and disk system.   Over time the energy released by the forming star, both dynamical and radiative, ablates the surrounding envelope, exposing the dense (n$_{\rm H_2}$ $\gg$ 10$^{5}$ cm$^{-3}$) disk to interstellar space.   The gas and dust rich protoplanetary disk then continues to evolve both physically and chemically until the gaseous disk dissipates and the system transitions to one dominated by large bodies and their surrounding debris (so-called debris disk systems).

A key facet of the disk evolution is the growth of the initially micron-sized dust particles to larger sizes, in which two aspects stand out. First, the gravitational settling of coagulating grains to the midplane removes dust from the surface layers of the disk \citep{wc_ppiii, dd04}.  This process is constrained via observations of the dust spectral energy distribution because dust settling decreases continuum emission in the mid- to far-infrared \citep{dalessio99,Chiang01}.  Surveys using the Spitzer Space Telescope infer dust depletion factors in the surface layers to be of the order of 100-1000 (relative to what is expected assuming interstellar grain abundances) in both Taurus and Ophiuchus \citep{furlan06, McClure10}.
Second, dust becomes radially stratified by size-dependent gas drag, a process known as radial drift \citep{whipple1972,wc_ppiii, Youdin13}.  
Small grains are coupled to gas motions while large km-sized planetesimals have significant inertia to resist the overall drag force.  In between, mm to tens of cm-sized grains \citep{Youdin13} radially drift inwards unless subject to a local pressure maximum generated by a variety of mechanisms \citep{Johansen07, Chiang10}.   There is now growing observational evidence for the pervasive presence of radial drift; one example is that the part of a disk composed of mm-sized grains seen in sub-mm emission is smaller in size than the part composed of smaller grains traced by scattered light or $^{12}$CO emission \citep{pdg07, isella07, panic09,andrews12}.   

\begin{turnpage}
\begin{figure*}[h!]
\begin{centering}
\includegraphics[width=1.3\textwidth]{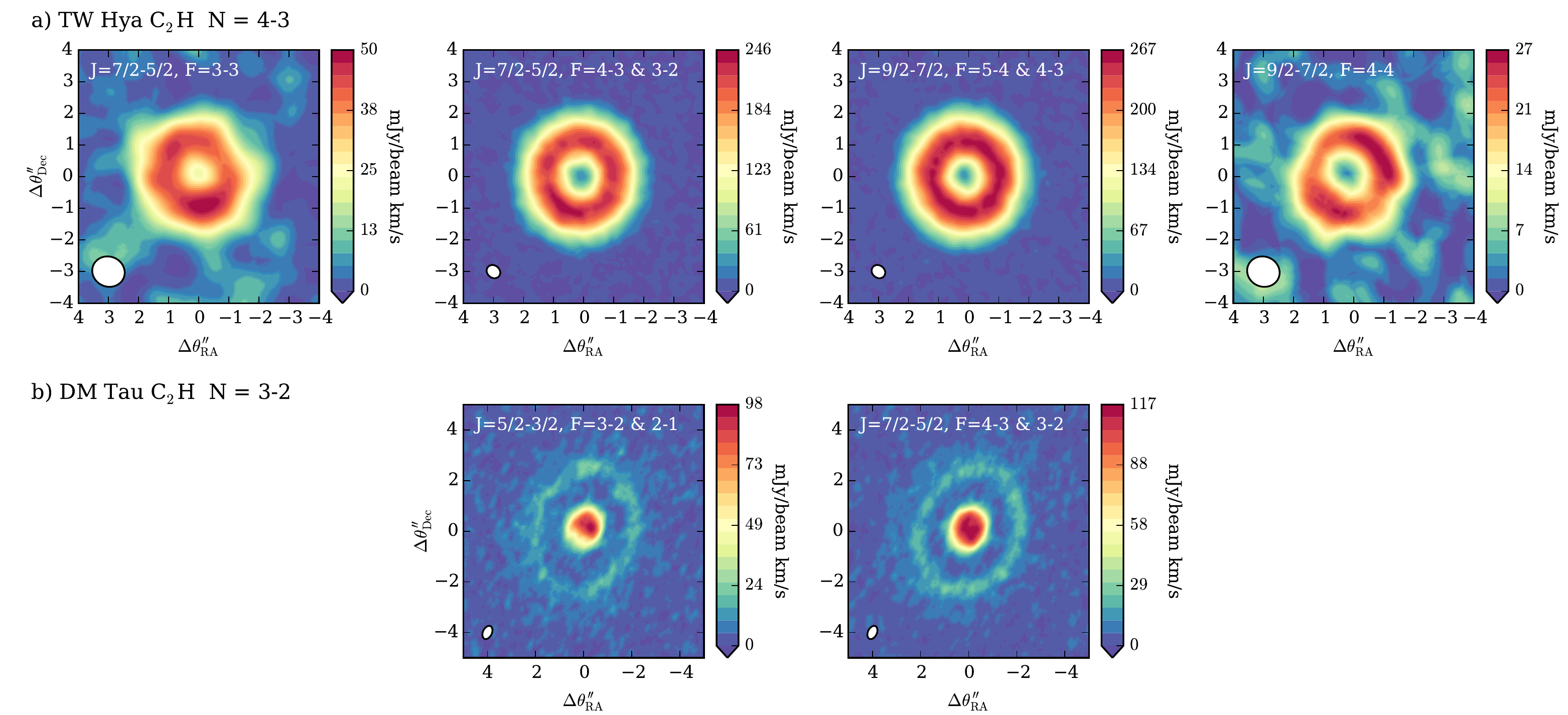}
\caption{Individual C$_2$H line detections for TW Hya (top row, $\rm N=4-3$) and DM~Tau (bottom row, $\rm N=3-2$). For both disks, the middle two columns are the sum of the emission of two partially blended hyperfine components as labeled. The weaker satellite components are not detected in the DM~Tau observations, consistent with the relative line strengths in the optically thin case. All molecular line emission maps presented in this figure and others are continuum subtracted.  \label{fig:indiv}}
\end{centering}
\end{figure*}
\end{turnpage}

What is less clear is how the evolution of dust affects the gas, and in particular the gas chemical composition.   The settling and growth of dust grains increases the transparency of the surface layers to UV radiation, while also increasing the local timescales for the freeze-out of the gas species \citep{aikawa99, an06,bb11a, Fogel11,Semenov11,Akimkin13}.  However, the effects are difficult to disentangle, and, while many chemical dependencies have been found in the models, strong correlations are yet to be discovered.  

More recently, there is observational evidence that the abundances of key volatile species
are depleted in the surface layers where the molecular emission originates --- the so-called ``warm molecular layer'' \citep{aikawa_vanz02}.
The generic expectation is that CO would be present as vapor with ISM abundance ($\sim 10^{-4}$ relative to H$_2$) in layers where the dust temperature exceeds $\sim$20~K. In gas below 20~K, low gas-phase CO abundances relative to the ISM are a hallmark of volatile freeze-out, and are seen clearly in observations of the cold regions of protoplanetary disks \citep{dgg97}. 
However, \citet{Dutrey03} and \citet{Chapillon08} find that the gas-phase abundance of CO may also be reduced in layers above the sublimation temperature.  A similar result is found for water vapor emitting from cold ($\sim 20-40$~K) layers well below the sublimation temperature \citep{bergin10b, hoger11a}. Still, molecular depletion is not the only explanation for the observed low volatile abundance in warm gas; the disk could also simply be less massive than the existing mass tracers predict.

The degeneracy between abundance and gas mass can be minimized by using a gas tracer, such as HD, to infer the H$_2$ mass \citep{bergin_hd}.
\citet{favre13a} use C$^{18}$O and HD emission to show that the CO abundance was significantly reduced in emissive layers in the TW Hya disk, while  \citet{Du15} find that both CO and water appear to be depleted from the surface beyond their respective snow lines.   They hypothesize, in concert with earlier and contemporary work\citep{Chapillon08, hoger11a, Kama16b, Kama16a}, that the ice-aided growth \citep{Gundlach15} and motions of grains deplete the upper layers of volatile species.  Taking into account this effect, \citet{Du15} are able to match the observed emission from \ion{C}{2}, \ion{C}{1}, CO, \ion{O}{1}, OH, and H$_2$O; they also predict that regions of the disk surface and outer radii will be preferentially depleted in oxygen and will become subsequently hydrocarbon-rich.  This composition arises because C- and N-bearing molecules exist in a sea of H$_2$ rich gas that is exposed to ionizing photons initiating hydrocarbon formation.
Contemporaneously \citet{Kastner15}, using the SMA, detected the presence of a bright C$_2$H emission ring in TW Hya.  

Here we present the spatially resolved distribution of C$_2$H and C$_3$H$_2$ emission in TW Hya and C$_2$H in DM Tau obtained with the Atacama Large Millimeter/submillimeter Array (ALMA).   Using the thermo-chemical model of \citet{Du14}, we demonstrate that hydrocarbon rings are a natural feature induced by the settling and drift of millimeter or larger sized grains, which removes UV opacity from the upper layers and outer disk, creating a UV-dominated disk (essentially a photon dominated region).  The chemical outcome of this drift is to deplete the upper layers of oxygen and raise the C/O ratio there.  This is also required to boost hydrocarbon emission in the ring to the detected levels.   In \S\ref{secObs} we describe the observational setup and in \S\ref{secHydrRing} we present images of the two sources.   In \S\ref{modeldesc} we use the thermo-chemical model of \citet{Du14} and \citet{Du15} to outline the ingredients required to produce hydrocarbon rings.  In \S\ref{discussion}, we present source-specific models that reproduce the observed emission, compare these predictions with an excitation analysis of TW Hya, and discuss potentials sources for the active carbon chemistry.  Finally, \S\ref{summary} summarizes these results and discusses the implications.

\section{Observations}
\label{secObs}

Observations of the DM~Tau and TW~Hya protoplanetary disk systems were acquired with the Atacama Large Millimeter/submillimeter Array (ALMA), under the project code ADS/JAO.ALMA\#2013.1.00198.S. DM~Tau was observed on June~6, 2015 with 37 antennas ($21.4-740.4$~meter baselines) and on December~25, 2014 with 38 antennas ($15.1-348.5$~meter baselines).  TW Hya was observed on December~31, 2014 and June~15, 2015 with 34 antennas ($15.1-348.5$~meter baselines) and 36 antennas ($21.36-783.55$~meter baselines), respectively. The total on-source integration time for DM~Tau in both configurations was 57.1~minutes. The total on-source integration time for TW~Hya was 43.4~minutes. For DM Tau, the nearby quasar J0510+180 was used for gain and absolute flux calibration. Quasar J0423-0120 was used for bandpass calibration. For TW Hya, the quasars J1256-057 and J1037-2934 were used for bandpass and gain calibration, respectively. Titan was used for the flux calibration.

\begin{deluxetable*}{lccc}
\tablecolumns{4}
\tablewidth{0pt}
\tablecaption{Hydrocarbon Observations \label{tab:obs}}
\tabletypesize{\footnotesize}
\tablehead{{Transitions} & Frequency & Peak Flux&Disk-Integrated Flux \\
                             & (GHz) & (Jy/beam) & (Jy km s$^{-1}$)}
\startdata
\\
{\em TW Hya } ($0\farcs48\times0\farcs39$) PA = $57.8^\circ$  & & &   \\
{C$_2$H }  & & &   \\
{~~}$\rm N=4-3, J=\sfrac{9}{2}-\sfrac{7}{2},~F=4-4$& 349.313 GHz & $0.660\pm0.007$ & $0.20\pm0.02$ \\
{~~}$\rm N=4-3, J=\sfrac{9}{2}-\sfrac{7}{2},~F=5-4$& 349.338 GHz &  $1.100\pm0.007$ & \multirow{2}{*}{$12.14\pm0.04^\dagger$ }  \\
{~~}$\rm N=4-3, J=\sfrac{9}{2}-\sfrac{7}{2},~F=4-3$& 349.339 GHz &  $1.002\pm0.007$ &  \\
{~~}$\rm N=4-3, J=\sfrac{7}{2}-\sfrac{5}{2},~F=4-3$& 349.399 GHz &  $0.862\pm0.007$ & \multirow{2}{*}{$10.72\pm0.04^\dagger$ }   \\
{~~}$\rm N=4-3, J=\sfrac{7}{2}-\sfrac{5}{2},~F=3-2$& 349.401 GHz &  $0.866\pm0.007$  &  \\
{~~}$\rm N=4-3, J=\sfrac{7}{2}-\sfrac{5}{2},~F=3-3$& 349.415 GHz & $0.089\pm0.007$ & $0.47\pm0.02$   \\
{{\em c}-C$_3$H$_2$ }$^\ddagger$  & & &   \\
{~~} $\rm (J_{K^+,K^-}$) $10_{1,10}-9_{0,9}$/$10_{0,10}-9_{1,9}$& 351.782 GHz &  $0.108\pm 0.007$  & $0.37\pm0.02$  \\
{~~~}$\rm (J_{K^+,K^-}$) $9_{1,8}-8_{2,7}$/$9_{2,8}-8_{1,7}$ & 351.966 GHz & $0.079\pm 0.007$ & $0.32\pm0.02$   \\
{~~~}$\rm (J_{K^+,K^-}$) $8_{3,6}-7_{2,5}$/$8_{2,6}-7_{3,5}$ & 352.194 GHz &  $0.067\pm 0.007$  &  $0.23\pm0.02$  \\
{~~~}$\rm (J_{K^+,K^-}$) $5_{5,1}-4_{4,0}$ & 338.204 GHz & $0.066\pm 0.007$  & $0.14\pm0.02$    \\
& &  & \\
{\em DM Tau} ($0\farcs56\times 0\farcs37$) PA = $-28.3^\circ$ & & &  \\
{~~}$\rm N=3-2, J=\sfrac{7}{2}-\sfrac{5}{2}, F=3-3$& 261.978 GHz &  $\lesssim 0.003$  &  ... \\
{~~}$\rm N=3-2, J=\sfrac{7}{2}-\sfrac{5}{2}, F=4-3$& 262.004 GHz &  $0.066\pm0.003$  & \multirow{2}{*}{$2.00\pm0.07^\dagger$}  \\
{~~}$\rm N=3-2, J=\sfrac{7}{2}-\sfrac{5}{2}, F=3-2$& 262.006 GHz & $0.063\pm0.003$ &   \\
{~~}$\rm N=3-2, J=\sfrac{5}{2}-\sfrac{3}{2}, F=3-2$& 262.065 GHz & $0.058\pm0.003$ & \multirow{2}{*}{$1.49\pm0.07^\dagger$}  \\
{~~}$\rm N=3-2, J=\sfrac{5}{2}-\sfrac{3}{2}, F=2-1$& 262.067 GHz &  $0.048\pm0.003$  &  \\
{~~}$\rm N=3-2, J=\sfrac{5}{2}-\sfrac{3}{2}, F=2-2$& 262.079 GHz &  $\lesssim 0.003$  & ...  \\
\enddata
\tablecomments{$\dagger$ The central hyperfine components are partially blended, so the reported disk-integrated fluxes are the sum of the indicated line pairs. $\ddagger$ In some cases detected lines are blends of degenerate transitions.} 
\end{deluxetable*}

The spectral setup targeted the C$_2$H $\rm N=3-2$ hyperfine complex in DM~Tau and C$_2$H $\rm N=4-3$ and multiple transitions of $c$-C$_3$H$_2$ in TW~Hya (see Table~\ref{tab:obs}).  
Observations of both sources were carried out with 122.07~kHz-width channels. From the pipeline calibrated measurement sets we self-calibrated the data using each source's bright continuum emission to improve the signal-to-noise. There was a dedicated continuum spectral window in the same sideband as the DM~Tau C$_2$H emission, which we use for the self-calibration. For the TW~Hya dataset, there was a pointing misalignment that may in part be due to TW Hya's high proper motion, and so we aligned the compact and extended data's phase-centers based on the continuum peak location. TW Hya's dedicated continuum window was in the opposite sideband from the C$_2$H lines, so we instead used the continuum data from the spectral window containing the lines, which were masked.  We imaged the measurement sets with the \texttt{clean} task in CASA version 4.3 using Briggs' weighting with a robust parameter of 0.5. For DM Tau, the resulting RMS noise level in the image is 2.5~mJy~beam$^{-1}$ in 0.279~km~s$^{-1}$, and for TW Hya is 7~mJy~beam$^{-1}$ in 0.15~km~s$^{-1}$ channels. The peak and disk integrated fluxes are reported in Table~\ref{tab:obs} with RMS uncertainties. An additional 15\% uncertainty should be applied for the absolute flux calibration uncertainty. The synthesized beam for the DM~Tau combined data sets has a shape and position angle (PA) of $0\farcs56\times 0\farcs37$ and $\rm PA=-28.3^\circ$, and for TW~Hya, $0\farcs48\times0\farcs39$ and $\rm PA = 57.8^\circ$. The moment 0 maps for the individual line detections for both sources are shown in Figure~\ref{fig:indiv}. 

\begin{figure*}[th!]
\begin{centering}
\includegraphics[width=0.65\textwidth]{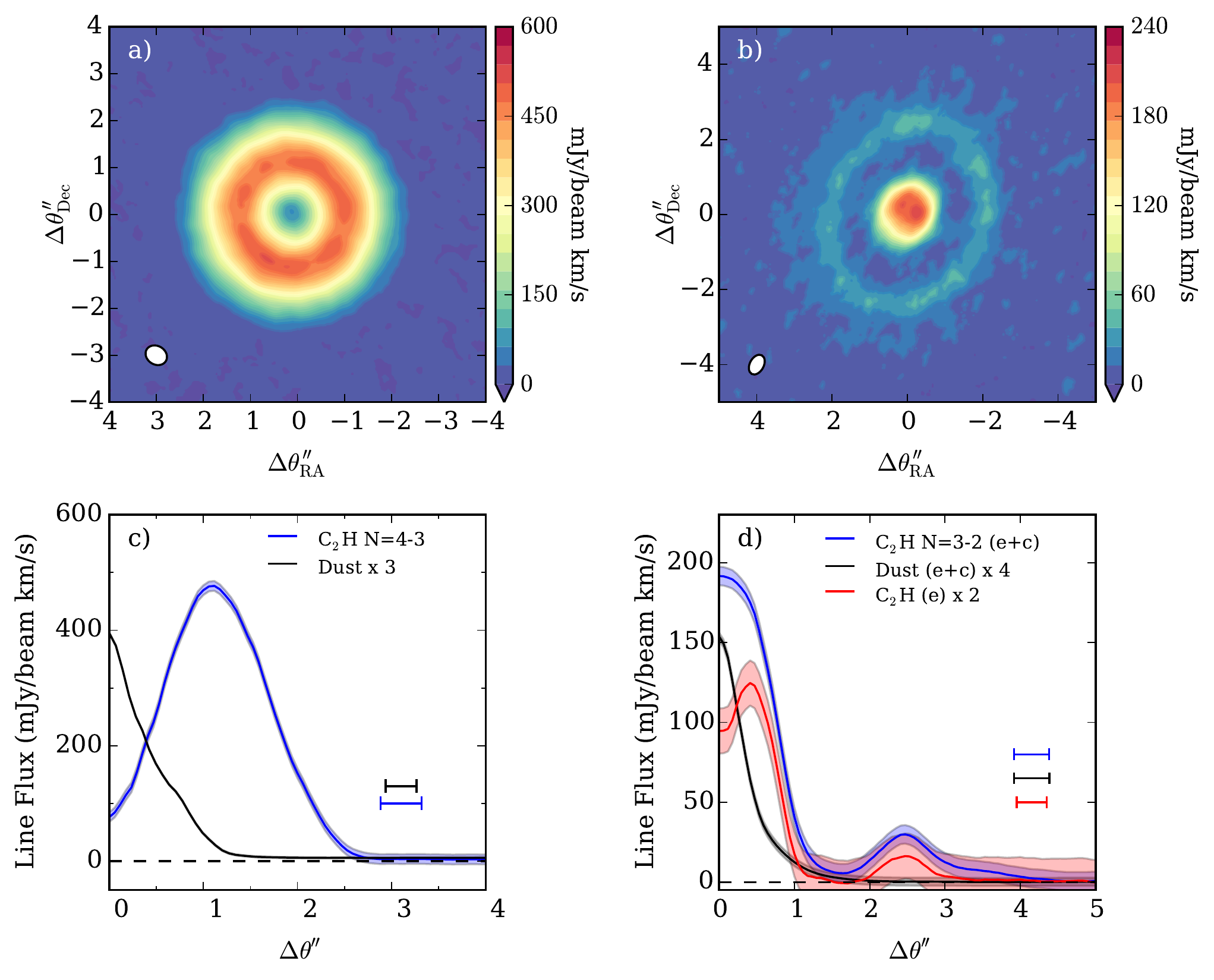} 
\caption{Top row: Stacked image of the individual C$_2$H detections presented in Figure~\ref{fig:indiv} for a) TW~Hya and b) DM~Tau. Bottom row: Deprojected line and continuum emission profile for c) TW~Hya and d) DM~Tau. The TW~Hya image profile is made from the compact TW~Hya data for the C$_2$H $\rm N=4-3$ (blue) and continuum (black). The DM~Tau emission profile is made from combined compact and extended data for the C$_2$H $\rm N=3-2$ line (blue) and continuum (black). The DM~Tau C$_2$H data re-imaged with higher resolution weighting and extended data only is shown in red. Continuum profiles are in units of mJy~beam$^{-1}$. \label{fig:stack}}
\end{centering}
\end{figure*}

We have stacked the individual lines to create a high signal-to-noise spectrally integrated map of the C$_2$H emission morphology, shown in Figure~\ref{fig:stack}.
For both disks, we derive a disk position angle and inclination from the outer C$_2$H ring, which we use to deproject the image to create a disk brightness profile.  The outer ring in DM~Tau has a position angle of $157\pm5^\circ$ and inclination of $i=36\pm3^\circ$.  The ring in TW~Hya has a position angle of $150\pm10^\circ$ and inclination of $i=7\pm4^\circ$.  The spectrally integrated line and continuum emission profiles are shown in Figure~\ref{fig:stack}, bottom row.

In the stacked profile, the inner C$_2$H emission in the DM~Tau disk flattens out towards the center.  This flattening motivated us to reimage DM~Tau using extended data only, with uniform weighting to optimize the image resolution.  The beam on the higher resolution images is $0\farcs30\times0\farcs48$, $\rm PA=-23.8^\circ$.  With this smaller beam, we successfully resolve a second ring in DM~Tau, shown in Figure~\ref{fig:hires}.  The RMS noise on the high resolution image is 6~mJy/beam in 0.279~km~s$^{-1}$ channels, and 8 mJy/beam~km~s$^{-1}$ on the spectrally integrated image.  We are unable to resolve the previously reported $R=20$~AU inner hole in the continuum \citep{Andrews11}, even with the higher resolution image.  However, detailed modeling of the continuum visibilities of the same data by \citet{Zhang16} shows evidence for a break in the inner disk on a similar scale.

\section{Hydrocarbon Rings}
\label{secHydrRing}

\subsection{C$_2$H Emission Distributions in TW Hya and DM Tau}

Both TW Hya and DM Tau are well studied sources.  
TW Hya is the closest young star with a disk at a distance of $54{\pm}6$ pc \citep{vanLeeuwen07}.
Its $^{12}$CO molecular distribution covers the full extent of the disk seen in scattered light \citep[][]{weinberger02, andrews12}.   
Based on a search for CO emission and \ion{Na}{1} absorption, the surrounding stellar association appears to be unassociated with its parent molecular cloud \citep{tachihara09}.  Furthermore, the extinction to this nearly face-on disk system is nearly zero, as demonstrated by the detection of almost its entire UV spectrum down to near the Lyman limit \citep{herczeg_twhya1}.
There is some controversy in the current literature regarding its spectral type and age, which is discussed in detail by \citet{Debes13} and \citet{vs11}, with the spectral range from K7 to M2 and an age of 8$\pm$4~Myr.   

DM Tau is an M1 star surrounded by a $\gtrsim$1000 AU molecular disk as seen in the
emission of $^{12}$CO \citep{sdg00, ddg03}.   It is located near the edge of the 
of the Taurus Molecular Cloud \citep{hbb01, pg_taurus}  at a distance of $\sim$140 pc
 \citep{Kenyon94, Schlafly14},
 where it may be partially obscured
by cloud material \citep[$A_V \sim 0.7^{\rm m}$;][]{espaillat2010}.   
Based on its spectral energy distribution and high resolution sub-mm dust continuum imaging, it is also a transition disk with a measured (sub-mm emission) inner gap radius of 19 AU  \citep{Calvet05, Andrews2011}.    The age of the system is estimated using models of pre-main sequence evolution with some uncertainty.   \citet{Hartmann98} placed DM Tau on the center of the age distribution in Taurus ($1-2$~Myr) at 1.5~Myr; but other work suggests it is slightly older \citep[$\sim 2 - 4.5$~Myr;][]{Kitamura02, Hueso05}.   At face value DM Tau appears younger than TW Hya.   This is based on the models of Hayashi track evolution discussed above, but also from the fact that molecular cloud material is present in close proximity of DM Tau, while absent near the TW Hya association.   

\begin{figure}[t!]
\begin{centering}
\includegraphics[width=0.35\textwidth]{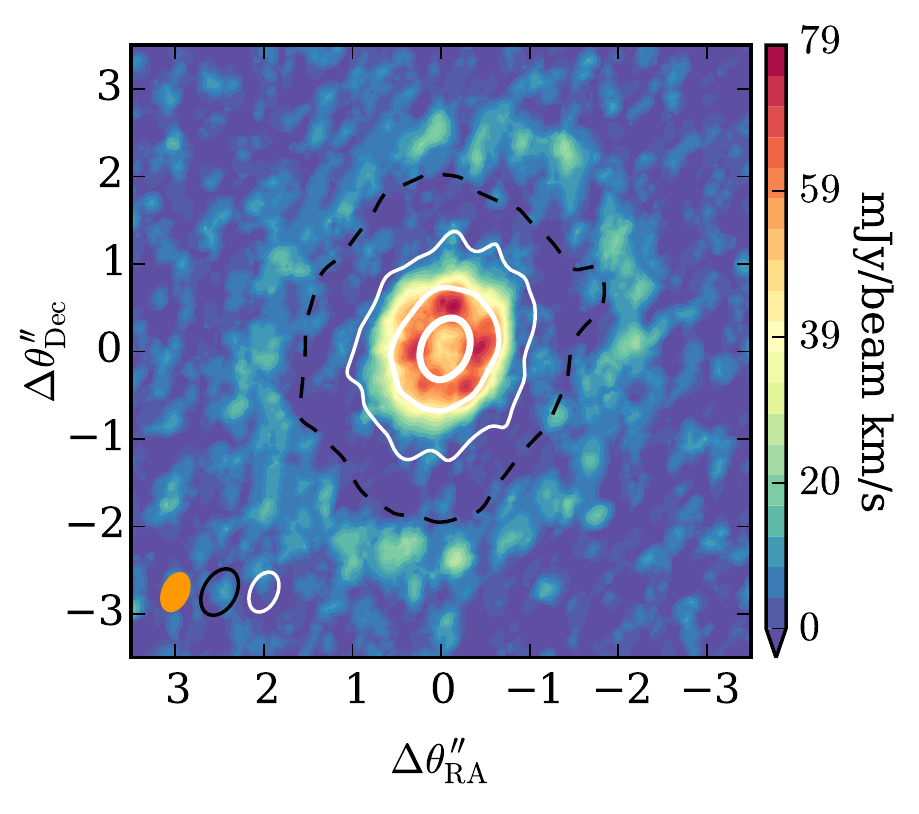} 
\caption{High resolution DM~Tau image using the extended data only, where the inner emission peak is resolved into a second ring. Overlaid are continuum contours, where the black dashed line shows the 3$\sigma$ continuum contour for the compact and extended data. The white contours highlight the 6, 24, and 96$\sigma$ continuum contours for the extended data only, going from thin to thick line weights. The inner dust emission is extremely steeply peaked at the location of the C$_2$H inner ring.   \label{fig:hires}}
\end{centering}
\end{figure}

Based on Fig.~\ref{fig:stack}, the dust emission at sub-mm wavelengths is confined with respect to the \ce{C2H} emission for TW Hya. The continuum emission at $\sim$850~$\mu$m originates from dust grains with sizes up to $\sim$millimeter \citep{Draine06}.  In TW Hya the \ctwoh\ emission peaks near the edge of the disk traced by sub-mm emission, down to our sensitivity limit.
A recent paper by \citet{Hogerheijde15} traces the edge of the surface density distribution at 820~$\mu$m to a radius of 47.1$\pm$0.2 AU ($0\farcs87$). The peak of the \ctwoh\ ring appears just beyond this edge at $R\sim 1\farcs1$, with a well resolved, deep central depression.   

 
In DM Tau there is significantly more structure in the \ctwoh\ emission.  In Fig.~\ref{fig:hires} we show the \ctwoh\ emission distribution using only the extended baselines. Overlaid on this image is the dust continuum emission with the black line showing the 3$\sigma$ dust emission edge at 1100 $\mu$m.  At the peak of the dust emission there is a central hole in the \ctwoh\  emission that encompasses the inner 100 AU (diameter).  This is followed by a ring of C$_2$H emission, also with a width ${\sim}100$~AU where the strongest emission is detected.   At much greater distances (R $\sim 350$~AU) there is a weaker and thinner C$_2$H emission ring. The fact that both TW Hya and DM Tau have hydrocarbon ring emission near the mm-dust edge is highly suggestive of a common process that might be present in evolving disk systems.  However, the large differences in the size scale of the rings, along with stronger central emission in DM Tau, requires detailed modeling to determine the import of specific mechanisms.

\subsection{$c$-C$_3$H$_2$ Ring in TW Hya}

\begin{figure}[b]
\begin{centering}
\includegraphics[width=0.35\textwidth]{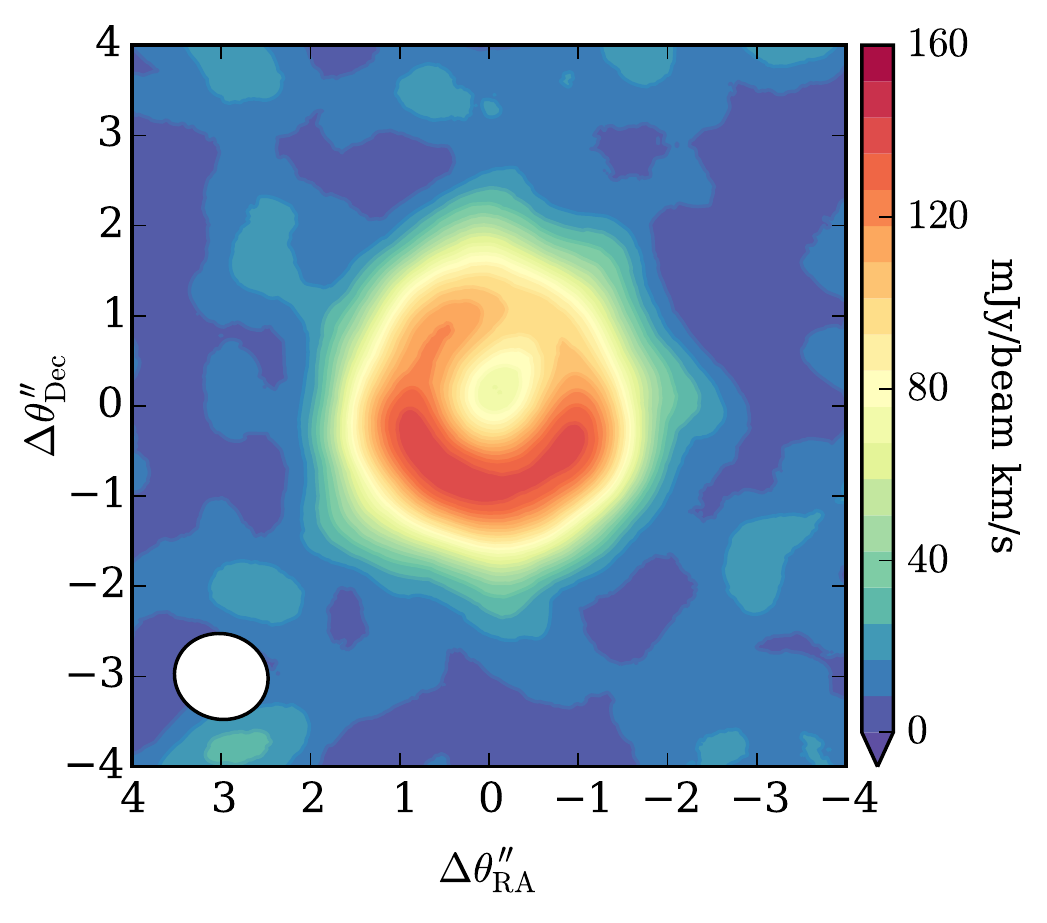} 
\caption{Stacked image of all detected transitions of $c$-C$_3$H$_2$ towards TW Hya. The transitions are listed in Table~\ref{tab:obs}.\label{fig:c3h2}}
\end{centering}
\end{figure}

Similar to \citet{Kastner15}, we mainly focus on the detection of C$_2$H rings in this paper.
In Fig.~\ref{fig:c3h2} we present stacked observations of $c$-C$_3$H$_2$ towards TW Hya that clearly show a ring with similar dimensions as C$_2$H. Thus, the detection of rings is not limited to C$_2$H alone. Instead we will argue below that it likely extends towards all {\em hydrocarbons}, as supported by the $c$-C$_3$H$_2$ detection in TW Hya.\footnote{The C$_3$H$_2$ data will be analyzed in a separate work (Cleeves et al, in prep.).}

\subsection{Dust Optical Depth and Ring Structure}
 
High dust optical depth is one mechanism that would produce inner holes in the emission of molecular tracers at these wavelengths.  For TW Hya, detailed modeling of the sub-mm continuum emission suggests the inner $\sim 40$~AU has $\tau > 0.5$ \citep{Hogerheijde16}.  Thus only the inner tens of AU could be optically thick.  In the high resolution study of \citet{Andrews16}, they discuss a possible transition to an optically thick disk near 20 AU; the modeling of \citet{Tsukagoshi16} also suggests that beyond 20~AU, the disk is optically thin at these wavelengths.  This is important as the C$_2$H emission begins to significantly decline inwards of 60 AU - where the dust is in these analysis is optically thin.   
  \citet{Kama16a} present another recent model of TW Hya and this model is also optically thin where C$_2$H emission begins to decline.   Thus high dust optical depth is not enough to explain the inner hole and a chemical effect is needed  \citep[see also][]{Kastner15}.


   DM Tau has evidence for two C$_2$H emission rings, an inner and an outer.   
Fig.~\ref{fig:hires} shows the contours of the strongly centrally peaked dust emission; nearly all the strongest dust emission ($> 6\sigma$) is confined within the main central bulge of the C$_2$H emission.    Here the drop of the C$_2$H emission towards the central hole is about a factor of two. 
 At face value this is consistent with what would be expected from the effects of high dust optical depth, as we would not detect the back side of the disk. Thus, at this resolution we cannot determine whether the central depression is a true chemical effect.   However, the outer ring appears at the edge of the dust continuum emission which must be optically thin at that radial distance.  The contrast between the C$_2$H peak at ($\sim 2\farcs4$) and trough ($\sim 1\farcs5$) is also $\ge$ 3.  Thus, the outer ring is not an effect of dust optical depth and is the result of a change in molecular abundance.

\begin{table*}
\centering
\caption{Parameters for the disk structure of \TWH{} and \DMT{}.  \label{tab:model}}
\begin{tabular}{@{} l c c c c c c c @{}}
\hline\hline
                &    $M_\text{dust}$     &   $r_\text{in}$   &   $r_\text{out}$   &  $r_\text{c}$  &   $\gamma$        &   $a_\text{min}$   &   $a_\text{max}$ \\
                &    ($M_\odot$)         &   (AU)            &   (AU)             &  (AU)          &                   &   ($\mu$m)         &   ($\mu$m)       \\
\hline
\multicolumn{7}{c}{\TWH{}}\\
Inner Disk       &    $1\times10^{-9}$    &   0.1             &   3.5             &  5             &   1.0         &   0.9                 &   2               \\
Sub-mm Disk      &    $2\times10^{-4}$    &   3.5             &   50              &  40            &   1.5         &   $5{\times}10^{-3}$  &   $10^3$          \\
UV-dominated Disk&    $1\times10^{-5}$    &   3.5             &   200             &  80            &   1.5         &   $5{\times}10^{-3}$  &   1               \\
\hline
\multicolumn{7}{c}{\DMT{}}\\
Inner Disk       &    $5\times10^{-7}$    &   1               &   4               &  5             &   1.0         &   $5{\times}10^{-3}$  &   $10^3$          \\
Sub-mm Disk      &    $1\times10^{-4}$    &   4               &   100             &  30            &   1.5         &   $5{\times}10^{-3}$  &   $10^3$          \\
UV-dominated Disk&    $4\times10^{-4}$    &   4               &   500             &  70           &   0.5         &   $5{\times}10^{-3}$  &   1               \\
\hline
\multicolumn{7}{l}{Note: The gas mass is 100 times the total dust mass.}\\
\end{tabular}
\label{tab:dust}
\end{table*}

\section{Physical and Chemical Models}\label{modeldesc}

In the following we will explore the physical/chemical structure within a realistic disk model framework to detail the origin of the ring structure and how to reproduce the hydrocarbon emission.     For this purpose we will use the disk model described by \citet{Du15} with basic parameters for \TWH{} and \DMT{} given in Table~\ref{tab:model}.  For the dust composition, we use an 8:2 mixture of astrosilicates and graphite.  Their optical constants are taken from \citet{dl84}\footnote{\url{https://www.astro.princeton.edu/~draine/dust/dust.diel.html}}.  The dust size distribution is a power-law with exponent 3.5 \citep{mrn}, and each disk component has different minimum and maximum grain sizes. A uniform gas-to-dust mass ratio of 100 is used over the whole disk such that $\rho_{gas}(r,z) = 100 \times \rho^{tot}_{dust}(r,z)$, regardless of the dust components.
We assume three components for both disks: an optically thin inner zone, a sub-mm intermediate one \citep{andrews12,Menu14, Hogerheijde16}, and a UV-dominated outer region.  The disk parameterization is the same as \citet{Andrews2009} (see also \citealt{LyndenBell1974,Hartmann1998}), namely, for each component, the surface density is
\begin{equation}
  \Sigma = \Sigma_\text{c} \left(\frac{r}{r_\text{c}}\right)^{-\gamma} \exp\left[-\left(\frac{r}{r_\text{c}}\right)^{2-\gamma}\right].
\end{equation} 
The volume density is calculated from the surface density based on vertical hydrostatic equilibrium, using temperatures calculated from a Monte Carlo radiative transfer treatment \citep{Dullemond2002}.
The UV radiation field uses the observed stellar spectrum of TW Hya \citep{herczeg_twhya1, herczeg_twhya2} as an input and is quantified in units of the wavelength integrated interstellar radiation field (ISRF) \citep[G$_0 = 1$ is equivalent to the UV field in the general ISRF;][]{habing68}.
\begin{figure*}[htbp]

\includegraphics[width=\linewidth]{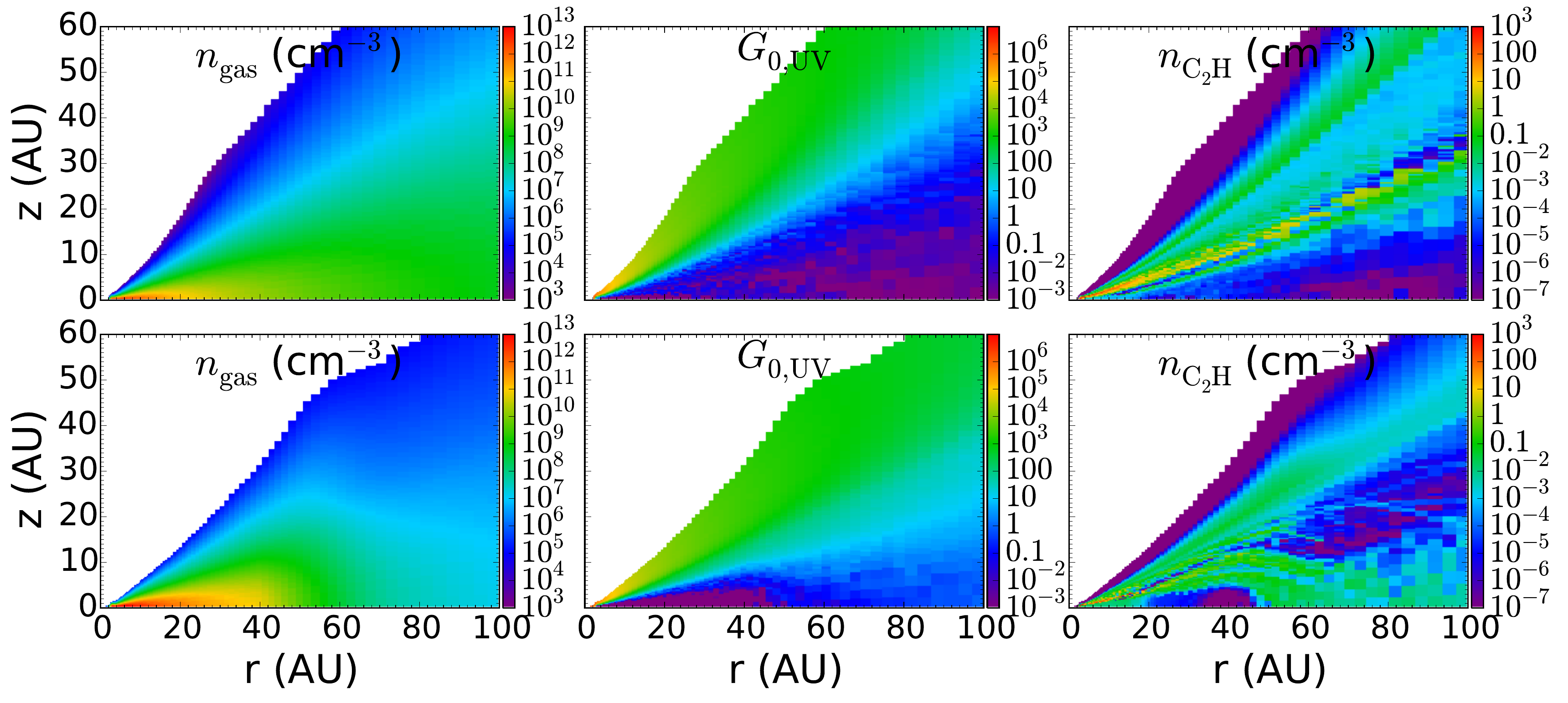}
\caption{Left column: The density structure of the smooth disk and the one with
transition; Middle: The UV field of the two cases; right: the \ce{C2H}
density distribution of the two cases.}
\label{figDistri}
\end{figure*}

The model for DM Tau is similar to TW Hya in that the dust optical constants and size distribution are the same, but the disk physical and chemical parameters are altered based on the observed dust spectral energy distribution, the size and overall distribution of the sub-mm emission disk from the data used here, and the C$_2$H molecular emission.  For our source specific modeling we did not adjust the intensity of the UV radiation field and we use the TW Hya radiation field for both systems. 
 In general our models show that the primary effects induced by the UV on the C$_2$H abundance occur with a scale much larger than caused by variations of a factor of a few, which is the known difference for these accreting stars \citep{bergin_h2, yang12, Schindhelm12}.  
 
 The X-ray luminosity for DM Tau is taken to be the preferred value from \citet{Henning10} of $L_X = 3 \times 10^{29}$~erg~s$^{-1}$, while for TW Hya we adopt $L_X = 1.6 \times 10^{30}$~erg~s$^{-1}$ \citep{Raassen09}.  Our group has explored the cosmic ray ionization rate within disks in general and for TW Hya specifically, suggesting that the rate may be significantly reduced due to interactions with the stellar wind  \citep{Cleeves13a, Cleeves15}.  Thus we have run our models with a rate that is reduced compared to the interstellar medium by two orders of magnitude ($\sim 10^{-19}$~s$^{-1}$, with an attenuation parameter of 96 g/cm$^2$).  Finally, our results are presented for models which have been sampled at 1 Myr for DM Tau; since TW Hya is known to be older we provide results from a model sampled at 5 Myr. 

Below we use the TW Hya model to explore some generic issues about the origin of C$_2$H ring emission; however,  it also provides important clues regarding the differing distribution observed in DM Tau.
We find that the overall structure of the {\em dust} disk is important in making the rings, but it is not enough by itself: some nontrivial chemical effect has to be incorporated to explain the observed morphology and intensity of the rings.  In \S~{5} we present  and discuss more detailed source-specific models for \TWH{} and \DMT{}.

\begin{figure*}[htbp]
\centering
\includegraphics[width=1.0\linewidth]{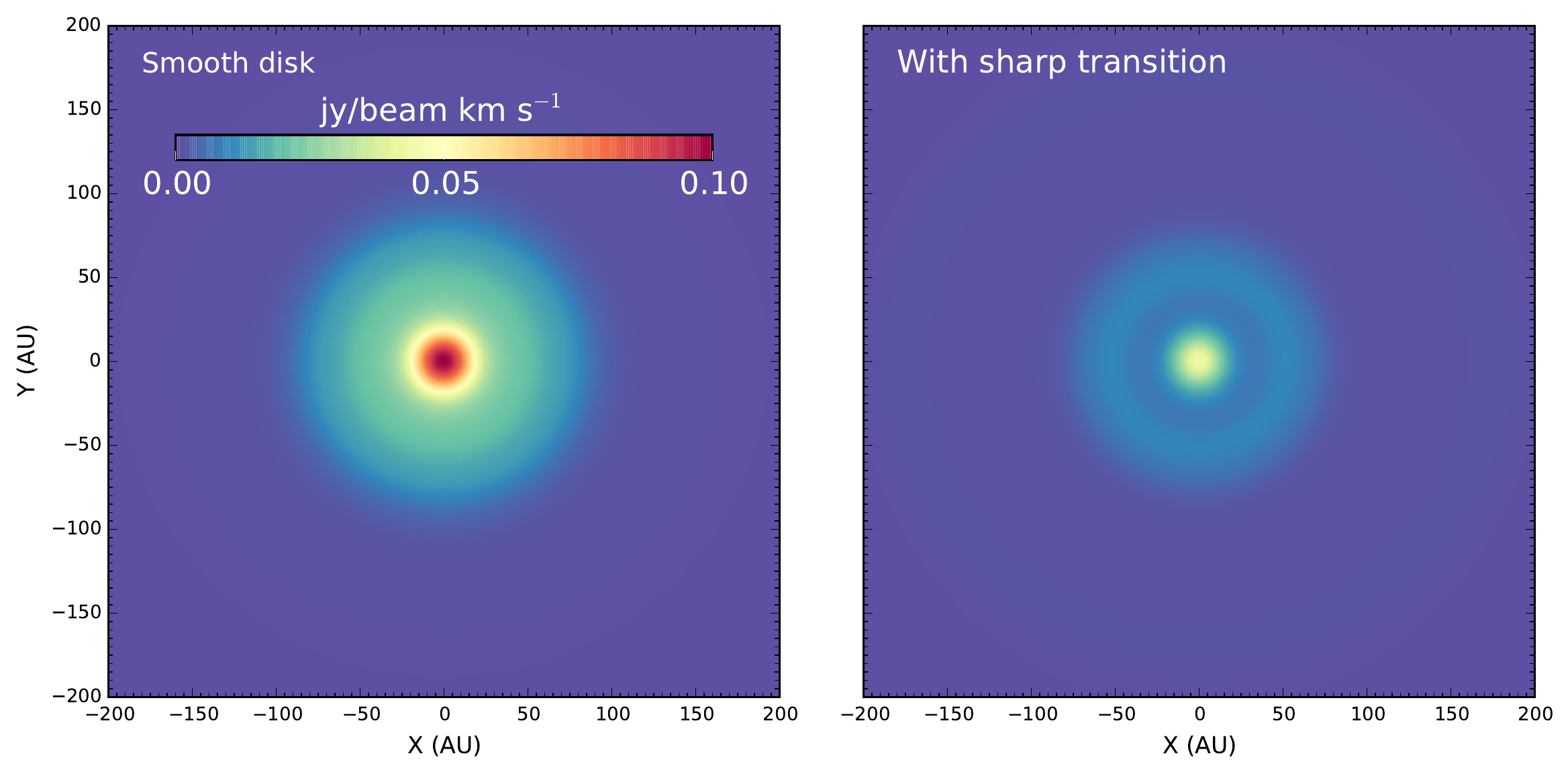}
\caption{The \ce{C2H} $J{=}7/2{-}5/2$, $F{=}4{-}3$ \& $3{-}2$ emission from a disk with smooth density profile (left
panel) and from a disk with a transition in column density (right); see \reffig{figDistri} for the two disk structures.
The images have been convolved with a Gaussian beam with full width half maximum of $0\farcs7$.  The
two panels share the same color scale.}
\label{figCMP}
\end{figure*}

\subsection{Ingredient \#1 for Hydrocarbon rings:  UV-dominated Surface and Outer Disk}
\label{secUV}

Models of dust evolution generally assume an interstellar grain size distribution as the starting point \citep[e.g.][]{Birnstiel12}.  The settling, coagulation, fragmentation, and radial drift of dust grains occur on timescales that depend on turbulence and the overall gas mass distribution of the disk.  The fact that the dust disk seen in sub-mm emission (Fig.~\ref{fig:stack}) is smaller than the gas-phase hydrocarbons and CO emission suggests that mm-sized (or larger) grains have undergone radial drift in both \TWH{} and \DMT{} \citep[see also the discussion by][]{Andrews15}.  These heavy grains are difficult for turbulence to loft to high altitude \citep{dd04}, which is supported by detailed studies of the dust emission for TW Hya \citep{andrews12}.  Thus, the larger grains are radially and vertically stratified.

Smaller grains are present since both systems show scattered near-IR light that extends beyond the disk traced by sub-mm emission \citep{weinberger02, Schneider05}.   Both disks have therefore evolved to the point where there is significant stratification in the dust mass.   To explore the effects of this solid mass stratification on the chemistry, we constructed two models: the first is the three-component model (inner disk, sub-mm disk, UV-dominated disk) as in \reftab{tab:dust} (similar to the eventual TW Hya model) which has 90\% of the dust mass confined within 50 AU.  The remainder of the dust mass is in smaller grains distributed over the whole disk.  This model therefore has a radial break in the dust distribution.  The second model assumes that all the dust mass is present as smaller grains that are smoothly distributed throughout the disk.

\reffig{figDistri} shows the density structure, the UV field (the $G_0$ factor), and the \ce{C2H} density for a disk with and without
dust mass stratification.   In the top panels, since UV-absorbing grains are present throughout the disk, the radiation field is strongest at the disk surface, and decreases with depth toward the disk midplane. In the case where most of the dust mass is confined, the overall density structure shows some differences, but the largest change is seen in the propagation of UV radiation.   Here the UV photons traverse the entire disk surface, wrapping around the mm disk, and are even present in the outer disk midplane ($G_0 \gtrsim$ 1).  In this template model (motivated by TW Hya) the dust mass difference between the mm-disk and the overall disk is $\sim$10.  Larger grains ($>$ mm-sized) have reduced ability to absorb UV radiation, with a factor of $\sim$20 reduction in the (mass) absorption coefficient at 1200 \AA\ between a 0.1~$\mu$m grain and a larger mm-sized pebble \citep[based on the optical constants from][]{dl84}.  The significant difference in mass and density between the large and small grain population in the stratified model overcomes this imbalance; it is the disk traced by mm-grains that is opaque to UV photons.   In the upper layers and outer disk, the gas is essentially a dense photon dominated region \citep[PDR;][]{th85}. For DM Tau, the UV dominated disk has more dust mass than the disk traced by sub-mm emission (\reftab{tab:dust}), but this is spread over a significantly larger volume.  In the following we will label this region the UV-dominated disk.

In the dense interstellar medium the reactive \ce{C2H} radical has been characterized as a tracer of transition zones
within PDRs where the gas is rich in atoms and simple molecules with short reactive lifetimes \citep[see][for a recent study]{Nagy15}.   Thus in \reffig{figDistri} there is a dramatic difference in the overall C$_2$H distribution.  In the smooth model C$_2$H is mostly confined to the upper layers of the disk where the UV field exceeds G$_0 > 1$.  In this model the primary formation pathways for C$_2$H are related to the overall C/O ratio in the disk, i.e. the presence of atomic carbon.  The key formation pathways are:
\begin{equation}
\begin{split}
\cee{C + H3+ -> CH3+}, \\
\cee{CH3+ + H2 -> CH5+}, \\
\cee{CH5+ + H2 -> CH3 + H2}, \\
\cee{CH3 + C -> C2H2}, \\
\cee{C2H2 + h$\nu$ -> C2H + H}, \\
\cee{C2H2 + H3+ -> C2H3+ + H2}, \\
\cee{C2H3+ + e- -> C2H + H2}. 
\end{split} 
\end{equation} 
We note that this list is not comprehensive and other reactions can play a role.  However, the critical link is the need for some atomic carbon to be extracted from CO via photodissociation.
Correspondingly, there is some structure in the C$_2$H abundance. The top layer traces the CO self-shielding zone and the lower vertical layer is the result of carbon extraction from CO via reactions with He$^+$, which is the product of He X-ray ionization. 

In the model with dust mass stratification, the lower optical depth for UV photons activates such chemistry over a greater disk volume.  Thus C$_2$H is observed to be present at high levels even in the outer disk midplane; the magnitude of the change is ${>}1000$ in abundance.  It is this difference that is one of the key pieces for the origin of the outer C$_2$H ring in both TW~Hya and DM~Tau.  \reffig{figCMP} shows resulting images of the \ce{C2H} emission in the N $= 4 - 3$ transition.  As can be seen, in the model with dust mass stratification, the \ce{C2H} emission exhibits a ring at $\sim$50~AU, close to the rim of the inner dust disk traced observationally by sub-mm emission.    It is near the outer radius of the pebble-dominated disk where the UV photons have their greatest effect -- as they can influence material closer to the midplane and extend the influence of the PDR over a greater column.  In this case pebbles refers to mm and cm sized-grains.
Thus, the smooth model does not show sharp features like this and dust evolution is expected to have a significant impact on the observed distribution of hydrocarbons \citep[see also][]{Semenov11}.

The emission distribution in the right-hand panel of \reffig{figCMP} is somewhat similar to that of DM Tau; although on a smaller physical scale.    The overall emission is also much weaker as this model is over an order of magnitude weaker than the \ce{C2H} ring emission in both sources.  An additional physical/chemical mechanism must also be operative to produce the central hydrocarbon emission hole detected in TW Hya.

\subsection{Ingredient \#2 for Hydrocarbon rings:  Depletion of Oxygen and Carbon, and Variable C/O Ratio}
\label{secOC}

Dust redistribution alone does not complete the picture as the stratified model does not match the level of observed emission; nor does it have an inner hole.
There are a few possibilities to raise the emission of \ce{C2H} in the ring.  First, there could be an order of magnitude more gas in the outer disk than included in our models.  But our models already assume a large gas mass of $\sim 0.02 - 0.04$~M$_\odot$.  In addition, our template model is based on the \citet{Du15} exploration of TW Hya and a higher outer disk gas mass will raise the HD emission to higher levels than observed \citep{bergin_hd}.  Based on HD, DM Tau has a comparable mass to that of TW Hya \citep{McClure16}.  Thus it is difficult for mass to be a primary variable both within and between a given source.   The fall off of mass with radius is also constrained by the radial distribution of small grains \citep[e.g.][]{Andrews15}.  Second, we could change the strength of the UV field, but again our models are constrained by the observed radiation fields of both sources, along with the modeled dust emission \citep{herczeg_twhya1, bergin_h2}.   Finally, the gas temperature could play a role, but is constrained by the emission from CO and its isotopologues \citep[][]{thi10, Du15}.    

\begin{figure}[htbp]
\centering
\includegraphics[width=0.8\linewidth]{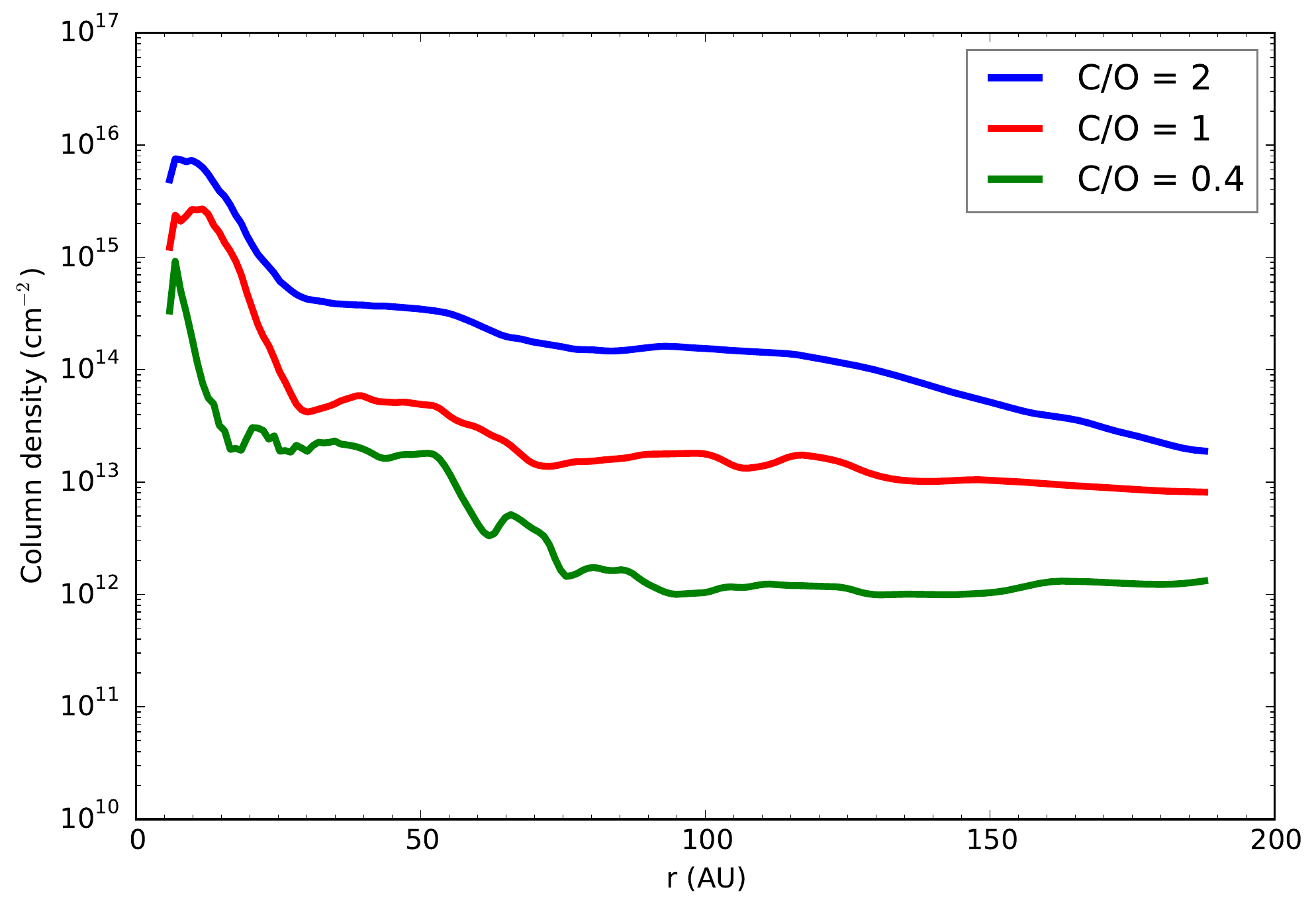}
\caption{Variation of the C$_2$H column density in the fiducial three-component model
of TW Hya (described in \S 4) for different (uniform) C/O elemental abundance ratio.}
\label{fig:Nc2h_co}
\end{figure}

We propose instead an alternative: that the dust mass stratification also implies a stratification in the {\em volatile inventory} of the disk.  Specifically, as the dust grains settle and drift, they also carry their ice mantles into the inner disk, leaving the upper layers of the outer disk depleted in volatile elements.  In the case of \TWH{} and \DMT{} this is supported by the weak emission from ground state transitions of water vapor \citep{bergin10b, hoger11a, Du15} along with the inferred depletion of carbon in TW Hya \citep{favre13a, Tsukagoshi15, Kama16a, Kama16b}.   Several papers have explored various aspects of this issue \citep{Bergin14, Furuya14, Reboussin15, Du15}.  A key point is that, due to its reduced volatility, water will be more susceptible to midplane sequestration than species such as CO.  Thus the upper layers of the outer disk will have C/O $\gtrsim1$ due to settling and drift.  This volatile stratification leaves C bearing species in a sea of H$_2$ with abundant ionization.  As predicted by \citet{Du15} this gas would be a hydrocarbon (and potentially also hydronitrogen) factory.   Indeed, in \TWH{}, as shown in Figs.~\ref{fig:indiv} and \ref{fig:c3h2}, the C$_2$H and $c$-C$_3$H$_2$ follow a common spatial distribution.
 
Here we will use our template model from \S~4.1 to explore the general effects of the relative carbon to oxygen depletion on C$_2$H, changing the overall C/O ratio uniformly in the entire disk.   In Table~\ref{tab:varco} we provide initial abundances of relevant species for different uniform C/O ratios in the disk.  Generally, and based on years of study of the interstellar medium, it is believed that the main carbon carrier is CO and carbonaceous grains, while the main carriers of oxygen are water ice and silicate grains, although other species contribute \citep[e.g. CO$_2$;][]{Whittet10}.   In the models below we ignore the refractory contributions and reduce the water ice abundance to mimic a changing C/O ratio.  
When the C/O ratio exceeds unity we must also redistribute some of the carbon out of CO into other carriers and in this instance we place the carbon in neutral atomic form.  

\begin{table}[htbp]
\centering
\caption{Initial Abundance Parameters for Variable C/O Ratio \label{tab:varco}}
\begin{tabular}{@{} l c c c@{}}
\hline\hline
Species  &   C/O=0.4 &    C/O = 1  &   C/O = 2  \\\hline
C        &   0       &    1.4 $\times 10^{-4}$  &   1.4 $\times 10^{-4}$  \\
CO       &  1.4 $\times 10^{-4}$ &   0         &  0\\
H$_2$O$_{\text{ice}}$     &   1.8 $\times 10^{-4}$ &    1.4 $\times 10^{-4}$  &   0.7 $\times 10^{-4}$\\\hline
\multicolumn{4}{l}{\footnotesize NOTE: Abundances do not reflect the total C/O ratio}\\
\multicolumn{4}{l}{\footnotesize as they do not include refractory carriers.}
\end{tabular}
\end{table}

\begin{figure*}[htbp]
\centering
\includegraphics[width=1.0\linewidth]{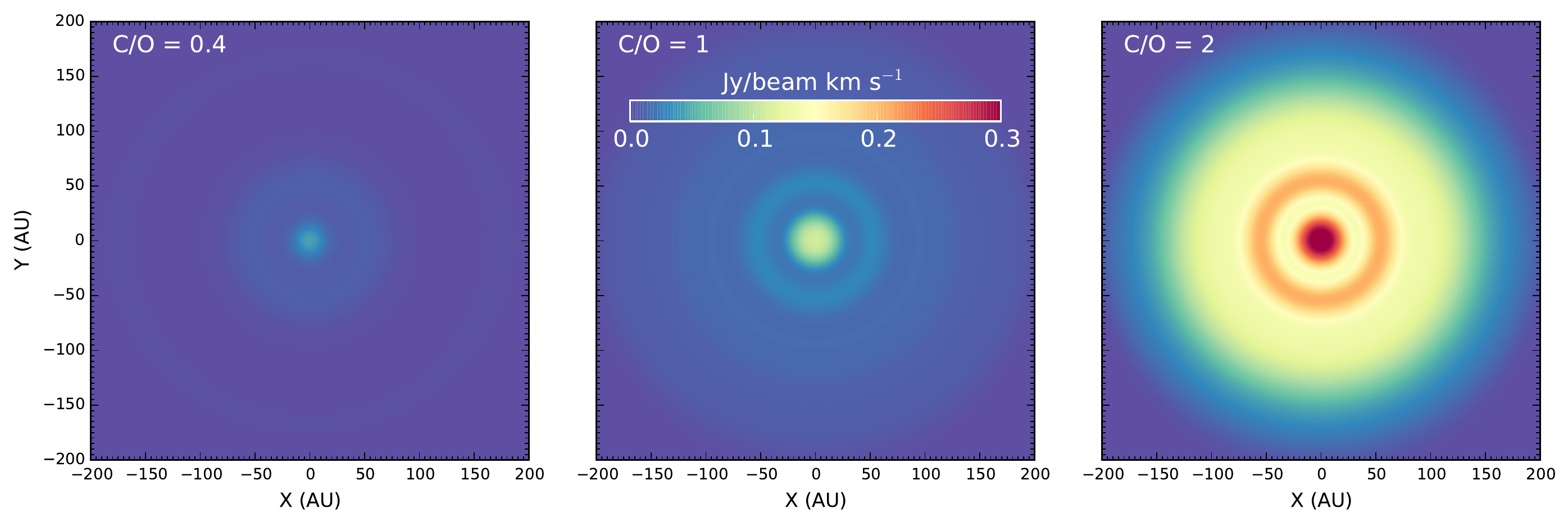}
\caption{Predicted combined emission for the two transitions $J{=}7/2{-}5/2$, $F{=}4{-}3$ \& $3{-}2$ of \ce{C2H} for models with different uniform elemental C/O ratio at the same integrated flux color stretch. For this model we use the three component TW Hya model described in \S 4.}
\label{fig:emit_co}
\end{figure*}

The results from these models in terms of implications for the radial column density 
variation of C$_2$H are shown in Fig.~\ref{fig:Nc2h_co}.  As one removes oxygen from the chemistry the C$_2$H column density increases, with a larger (one order of magnitude) increase beyond $\sim$75 AU.  The key issue here is the role oxygen plays in the destruction of carbon-bearing species that are not as stable as CO or CO$_2$.   In the case of the radical C$_2$H the main destruction paths are as follows: 
\begin{equation}
\begin{split}
\cee{O + C2H -> CO + CH}, \\
\cee{N + C2H -> C2N + H}, \\
\cee{C + C2H -> C3 + H}, \\
\cee{C+ + C2H -> C3+ + H}.
\end{split} 
\label{eqC2Hdestroy}
\end{equation} 
Atomic oxygen is present in the model due to photodesorption of water ice and its subsequent photodissociation \citep{hkbm09}.    But if oxygen is sequestered on larger grains that settle to the midplane and drift inwards, then the oxygen abundance of the upper layers and of the outer disk will be reduced.  The lack of oxygen removes a key species that is keeping the hydrocarbon chemistry in check, but also elevates abundances of other important species such as N, C, and C$^+$. This process elevates the abundance of hydrocarbons as discussed by \citet{Du15}.
Modest removal of oxygen produces relatively large enhancements in the C$_2$H column density across all radii.

In Fig.~\ref{fig:emit_co} we show the resulting emission profiles for models with different C/O ratios.  These simulated images show some structure, but here we focus solely on the overall intensity.   The underlying physical model was originally developed for \TWH{}, so one can compare with the panel for the $J{=}7/2{-}5/2$, $F{=}4{-}3$ and $3 {-}2$ combined transitions in \reffig{fig:indiv}.  As can be seen for the model with a ``normal'' abundance of carbon and oxygen (C/O = 0.4) the emission is well below observed values.  A C/O ratio of unity comes closer, but C/O $\gtrsim$ 1 is needed (in this model) to approach observed values.   These models roughly match the distribution seen in DM~Tau; however, this model is at a much smaller physical scale.  Nonetheless it does hint that this process could be ongoing in that source.

The uniformly increased C/O ratio does not explain the inner hole of the \ce{C2H} emission for TW~Hya.  There are two possible ways to create the inner hole.  The first is to enhance the oxygen abundance relative to carbon in the inner disk, which is exactly the opposite of the outer disk.
The spatial C/O ratio can be influenced by the various snow lines, which is discussed in detail by \citet{omb11}.
The size of the inner hole is $\sim$20 AU, which lies inside
the surface CO snow-line of \TWH{} at $\sim 25 - 30$~AU \citep{Qi13a}.
When CO ice sublimates it can influence hydrocarbon chemistry.  However, this is a two way street in the sense that when dissociated, CO provides an oxygen atom that readily destroys C$_2$H (going back to CO), it also leaves a reactive C$^+$ ion in the gas, which can produce hydrocarbons \citep[e.g.][]{Bergin14}.  The H$_2$O sublimation front has been studied by \citet{Blevins16} and are inferred to exist within the inner few AU of the disk surface for T~Tauri disk systems.  CO$_2$ is more difficult to study but is known to be present in ices \citep{Oberg11c} and comets \citep{mc11}; based on estimated bond strengths it should evaporate between the snow-lines of water and CO \citep{Martin-Domenench14} in between $\sim5-10$~AU.  The injection of \ce{H2O} and/or \ce{CO2} will lead to a minimum hole size in the C$_2$H abundance.

However, although the enhancement of oxygen relative to carbon could create an inner hole, this leads to an inconsistency with observations if carbon itself is at the ISM abundance.  It is known from observations of CO isotopologues that the CO abundance is reduced relative to the canonical ISM value of $10^{-4}$ by nearly two orders of magnitude \citep{favre13a,Nomura16, Schwarz16}.  This forces us to the second alternative that the abundance of chemically-active carbon must be reduced in the inner $\sim$25 AU, while the behavior of oxygen in the inner disk is unclear. The strong water emission in TW Hya \citep{Zhang13} shows that there is some sublimation within the inner 10 AU.    It is also plausible that some, potentially large, fraction of the carbon atoms are assimilated into refractory organic grains and do not return to the gas phase even in the inner warm region.

In summary, the settling and migration of icy grains reduces the UV opacity in the surface and the outer disk.  At the same time, the grains initially carry greater amounts of water than carbon-bearing species, leading to a C/O ratio of at least $1$ in the upper layers of the outer disk.   A variety of mechanisms have been explored in the literature that liberate carbon from CO \citep{Bergin14, Furuya14, Reboussin15, Du15}, which can enhance the C/O ratio to higher levels and increase the abundance of C$_2$H and other hydrocarbons.    Over time this can also deplete the inner disk of carbon.  When combined with sublimation closer to the star a deep central hole can develop over time.

\begin{figure*}[htbp]
\includegraphics[width=\linewidth]{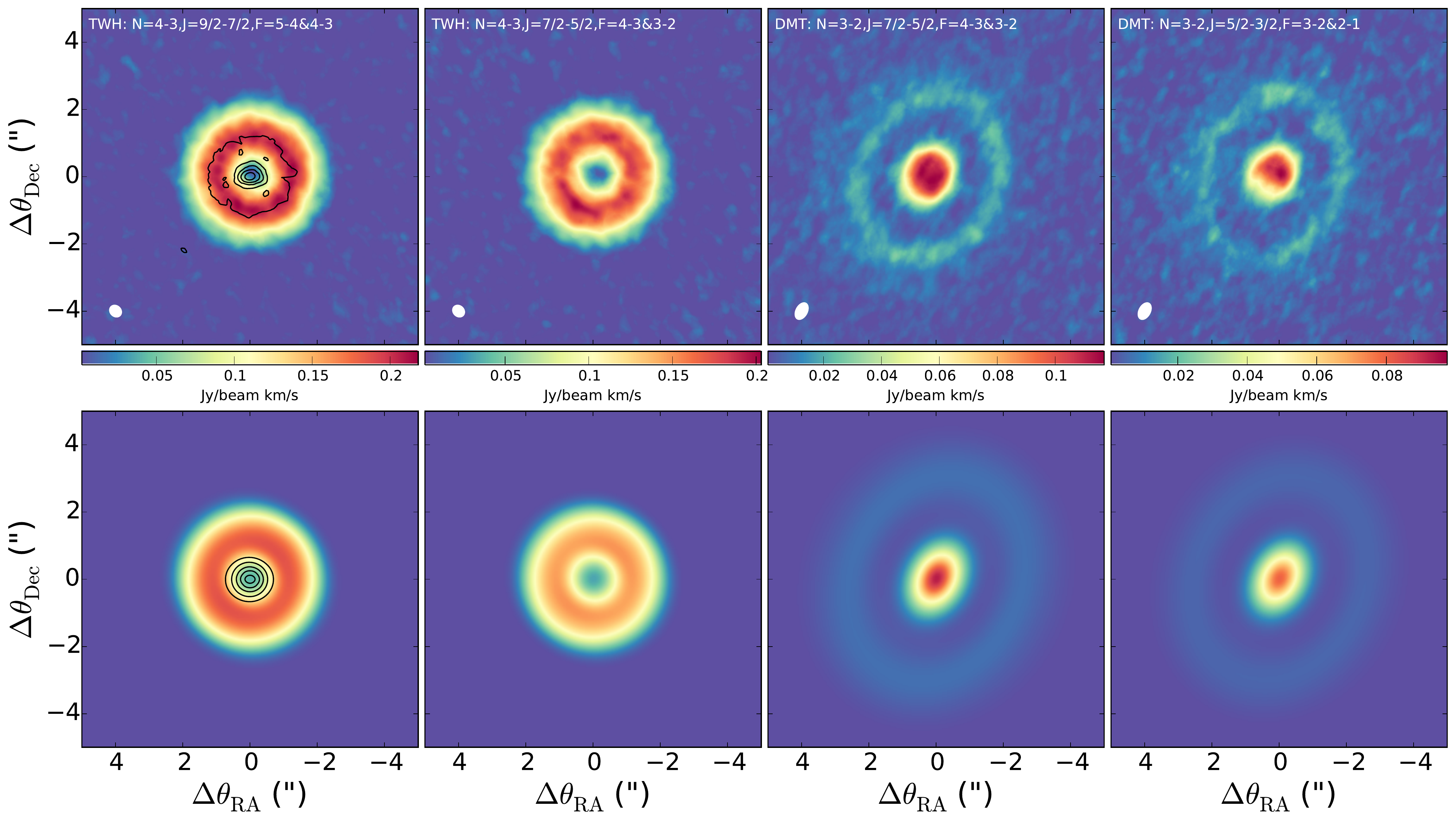}
\caption{The best fit model (bottom panels) showing the \ce{C2H} emission of \TWH{} and \DMT{} compared
to the observational data (top panels).  The modeled emission have been convolved with the beam of the observation.  Contours overlaid on the top-left and bottom-left panel are the observed (see \citealt{Schwarz16}) and predicted \ce{C$^{18}$O} ($3-2$) emission.}
\label{fig:best}
\end{figure*}

\section{Discussion}\label{discussion}

There are a number of issues unresolved.  \TWH{} and \DMT{}  disks have very different observed emission morphologies.  \DMT{} has stronger central emission, though its central region seems to be a resolved hole or flattening (see \reffig{fig:hires}), with a much more extended, weaker, outer ring.  The \DMT{} disk is substantially larger in size, although the gas masses of the two are comparable \citep{McClure16}.   Furthermore, we have assumed that CO is the source of carbon for C$_2$H, but the photodegradation of aliphatic or aromatic hydrocarbon grains could also be a source term for the needed carbon \citep[see discussion in][]{Kastner15}.   In the following we present source specific models to match the emission for both sources, explore the excitation characteristics of C$_2$H emission, and discuss additional sources of carbon for the production of the observed species.

\begin{figure*}[htbp]
\includegraphics[width=\linewidth]{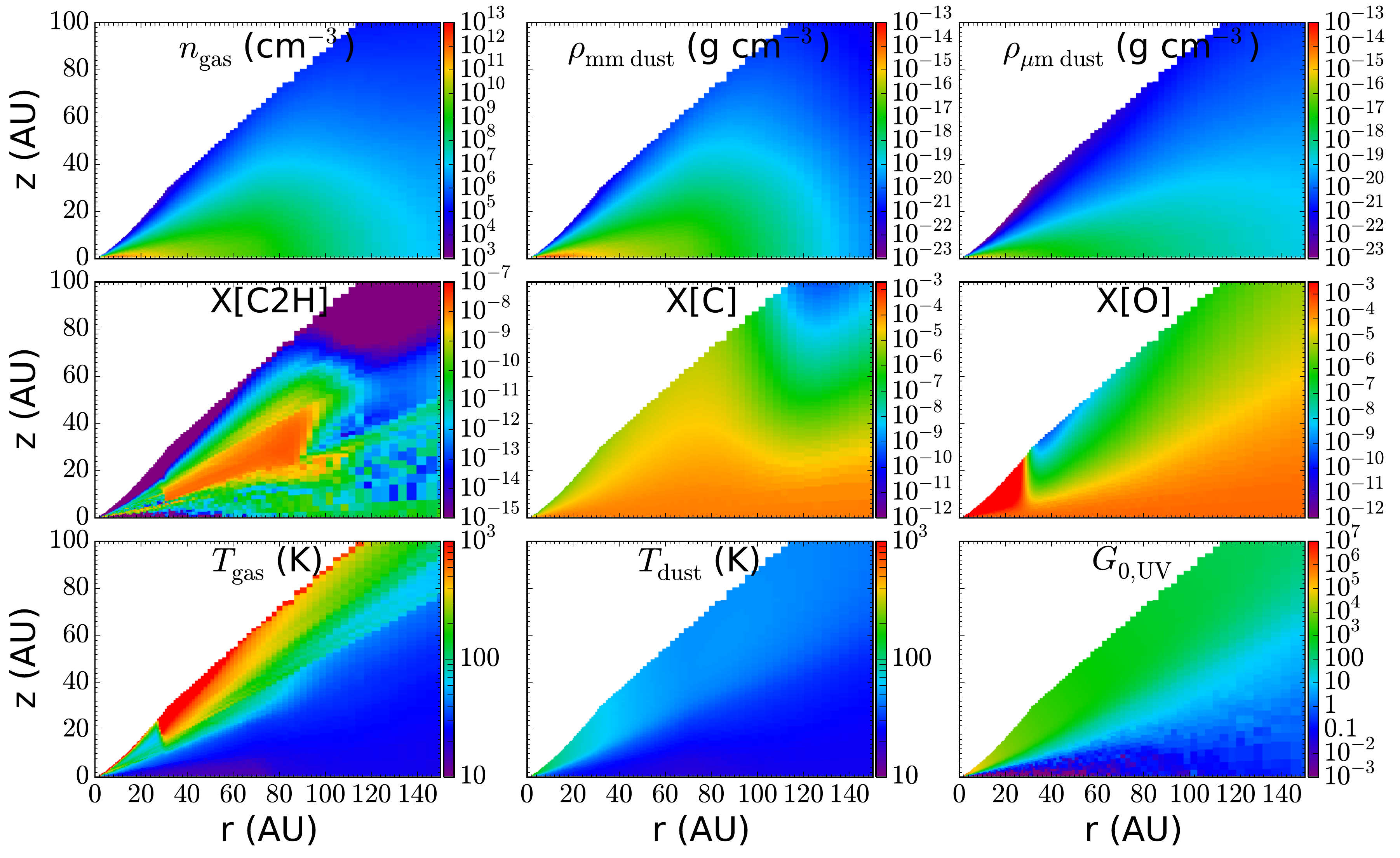}
\caption{Distribution of physical and chemical parameters in the ``best-fit'' model of TW Hya.   The upper panels provide the hydrogen nuclei number density (in cm$^{-3}$) and large grain and small grain density (in g~cm$^{-3}$).  The middle panels show the C$_2$H abundance relative to hydrogen nuclei, and the elemental abundances of carbon and oxygen.  The bottom panels show the calculated gas and dust temperatures followed by the UV radiation field in units of G$_0$. }
\label{fig:twh}
\end{figure*}

\subsection{Source Specific Models}
\label{sec:mod}

In our list of key ingredients to produce the hydrocarbon ring emission we illustrated  the increased penetration of UV radiation due the combined effects of settling and radial drift and showed how \ce{C2H} emission can be boosted by global changes in the C/O ratio.    To link these physical and chemical effects together, the distribution of key carriers in the elemental abundance pools must be non-uniform.  The ice mass likely follows the dust mass; hence the relative distribution of the volatiles left in the gas must reflect this in some fashion.  Furthermore there might be some reset process interior to the snow lines of key carriers, though this is not guaranteed to happen for two reasons.  First, some fraction of the ice may be sequestered into large km-sized planetesimals and will not evaporate at the traditional snow-line \citep{piso15}.  Second, if CO processing is very efficient such that the oxygen from CO is incorporated into water and the carbon into larger species \citep[e.g.,][]{Bergin14}, then the CO abundance would be depleted in the gas and at the surface CO snow line.  Thus on crossing the snow line there will be net loss of oxygen and carbon in the gas phase, the exact amount of which depends on the timescales for chemical processing.

In the appendix we provide the parameterization we adopt to model the uneven distribution of elements.
This parameterization is somewhat arbitrary and more realistic chemical models need to be constructed that include the formation of bodies large enough to stop drift and halt the formation of small grains via collisions that can be lofted to the warmer UV-exposed surface, where volatiles can be released from the ice mantles.   However, our  parameterized model captures key aspects of the proposed picture where volatiles are sequestered in the midplane and, due to drift, reside predominantly as ices in the inner tens of AU of the disk.

The ``best-fit'' emission images of \ce{C2H} for both disks are shown in \reffig{fig:best}. The parameters describing the basic disk structure are shown in \reftab{tab:model} and the corresponding parameters for the elemental abundance distribution of carbon and oxygen are provided in \reftab{tabDepl} for \TWH{} and \DMT{}, the meanings of which are defined in appendix~A.

\begin{table}[htbp]
\centering
\caption{Depletion parameters for oxygen and carbon in \TWH{} and \DMT{}, as defined in equations (\ref{eqdep1}--\ref{eqdep3}).}
\begin{tabular}{ccccc}
\hline
\hline
        &   $r_0$   &   $r_\text{s}$    &   $a$     &   $b$     \\
        &   (AU)    &   (AU)            &           &           \\
\hline
TW Hya (O)  &   15      &   3               &   0.9     &   0.6     \\
TW Hya (C)  &   60      &   5               &   0.7     &   0.2     \\
DM Tau (O)  &   70      &   20              &   1       &   0.2     \\
DM Tau (C)  &   90      &   20              &   0.9     &   0.6     \\
\hline
\end{tabular}
\label{tabDepl}
\end{table}

\subsubsection{TW Hya}
\label{sec:twh}

\begin{figure*}[htbp]
\includegraphics[width=\linewidth]{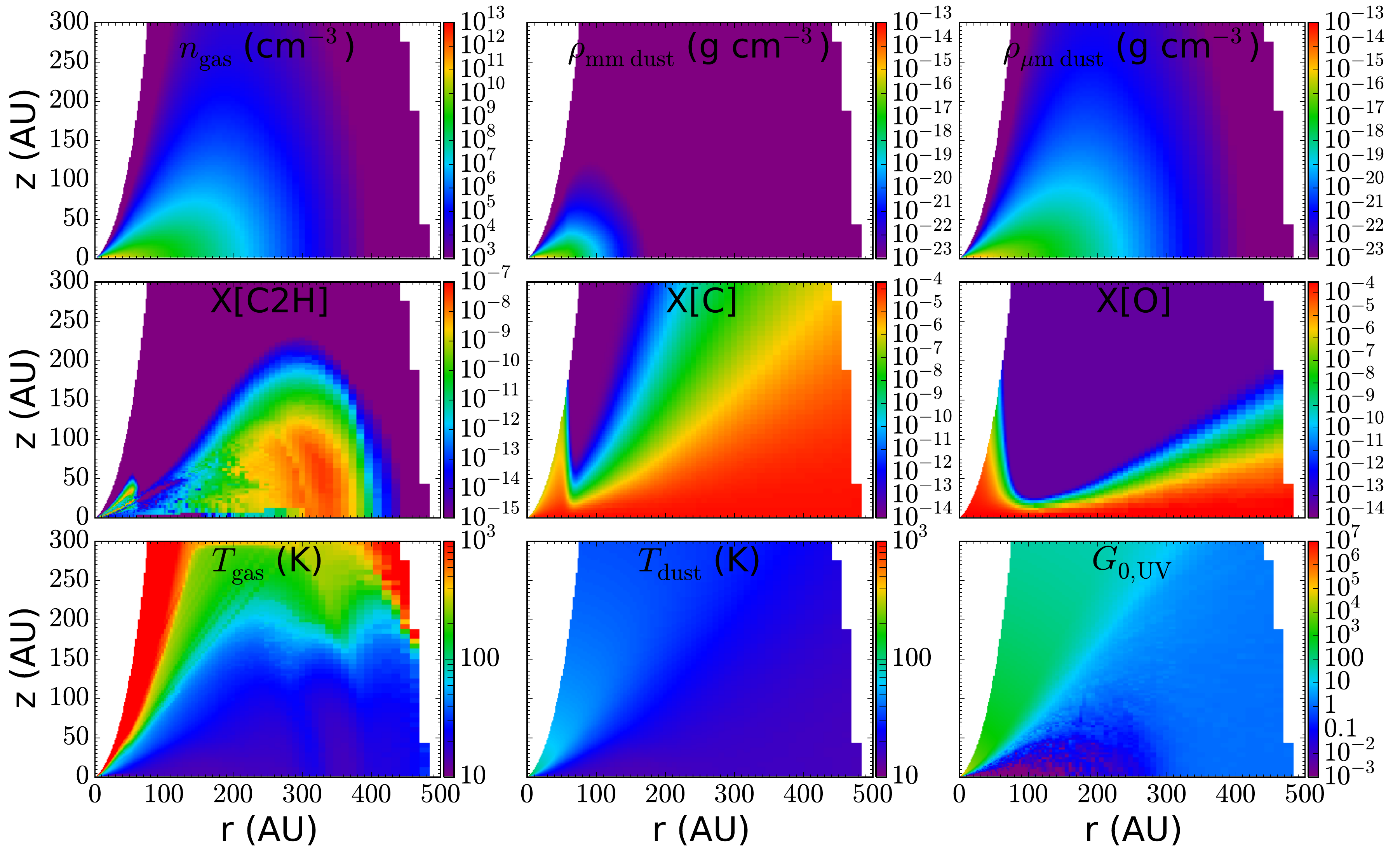}
\caption{The same as Fig.~\ref{fig:twh}, but for DM~Tau.}
\label{fig:dmt}
\end{figure*}

In Fig.~\ref{fig:twh} we show the distribution of physical parameters for the ``best fit'' model for the distribution of C$_2$H in TW Hya. For reference we show the contribution function for this emission map in Appendix B (Fig.~B.1).   In this model there is one central facet and two areas of uncertainty.
The central requirement is that in the ring, the C/O ratio is $\gtrsim$1 to reproduce the observed intensity.   C$_2$H is formed in the UV-dominated disk in the upper layers and down to the midplane wrapping around the inner disk comprised of larger grains.
The two areas of uncertainty regard:  (1) the central hole in the emission and (2) the outer radius of the emission ring, which is beyond primary edge of the pebble disk but interior to the outer radius of $^{12}$CO emission.   With our assumptions discussed below, the predicted emission distribution shown in Fig.~\ref{fig:best} matches observations within a factor of 1.5.

For the inner edge, in our models the elemental carbon abundance is reduced in the upper emissive layers within 20~AU.  This is required by both the CO isotopologue data \citep{favre13a,Nomura16,Schwarz16} and the \ce{C2H} data.
One possible reason for this may be
an uneven distribution of pebbles in the disk. \citet{Tsukagoshi16}
 find that the TW Hya mm-wave (138 and 230 GHz) dust spectral index increases from the inner edge until $\sim 20$~AU.  This is generally interpreted as due to grain growth in regions with lower values of the spectral index \citep[e.g.][]{Perez15}.   Thus perhaps there is an excess of  larger cm-sized or larger grains that incorporate carbon in more refractory form giving rise to the central depletion.
  The modeled \ce{C$^{18}$O} emission also matches the result of \citet{Schwarz16}, which is overlaid on the top/bottom left panels of Fig.~\ref{fig:best}.  A normal ISM abundance of carbon would over-predict the \ce{C$^{18}$O} emission, and would not be able to reproduce the high contrast (i.e. the ratio between the intensity in the ring to the intensity in the central hole) of the \ce{C2H} ring.  


The second issue then regards the physics/chemistry dictating the position of the outer ring edge.   From observations
of $^{13}$CO in the TW Hya disk \citep{Nomura16, Schwarz16} the emission intensity exhibits a sharp decline near 60 -- 70~AU ($\sim 1.2''$) at a point where the C$_2$H emission also begins to decline.   Thus, one possibility is that the edge of the emission ring is set by an overall decrease in the {\em gas} surface density.  Of course $^{12}$CO and scattered star light are observed outside of $\sim$2$''$ \citep{andrews12, Debes13}, but this could be due to a more tenuous outer disk.   Alternately, the outer edge could also be set by carbon depletion (which is the case in the parameterization of \reftab{tabDepl}) and the hydrocarbon emission rings are just maxima where carbon exists in the gaseous state.  If the temperature of the young TW Hya disk was below the CO sublimation temperature ($\sim 20$~K) grains that grew in the midplane at large distances and drifted inwards would ``steal'' carbon from the outer disk leaving a carbon depleted warm molecular layer on the surface as the carbon repository. 
Thus if any part of the disk will suffer strongly from these effects it must be the outer disk where the combined effects of settling and drift deplete this region of icy grains.  Since drift appears to be a general result, we would predict similar chemical effects in all disks where evidence of drift is observed.

In our models we have made an assumption of a constant gas/dust mass ratio throughout the disk, regardless of our assumptions regarding the {\em grain} mass distribution. Consequently, in our generic model where the grain mass is concentrated in the inner tens of AU and in the midplane, the gas is also centrally concentrated and simultaneously depleted in the outer disk. The best estimate we have on the gas mass comes from HD, which is a better probe of the mass of the inner disk \citep[$<$ 40 AU;][]{bergin_hd}.
Therefore, the gas mass of the outer disk is less well constrained. High resolution studies of CO isotopologues towards TW Hya with ALMA find that that the $^{13}$CO emission begins to sharply drop beyond 60 AU to below detectable levels \citep{Nomura16, Schwarz16}. This drop might hint at a change in the outer disk gas mass (or to the carbon abundance).
Thus, it is fair to question whether the reproduction of the C$_2$H emission requires the a priori variation in the C/O ratio to boost its abundance, as the mass could be higher in the outer disk than what is currently in our models. 

To check the effect of the outer disk gas mass, we have run a model with a factor of 10 higher gas mass.  We find that the resulting C$^{18}$O emission is significantly elevated over observed levels, while the C$_2$H emission is still well below observed values. To reproduce the observed intensities in this higher mass model, we require carbon depletion and C/O $>$ 1. There are additional constraints for different disk gas models such as \citet{Kama16a} who explore the abundance of C$_2$H (and other carbon bearing atoms/molecules) in TW Hya assuming a gas/dust ratio of 200.   They find a similar result and require carbon depletion and C/O $>$ 1.  Thus the high C/O required to explain the C$_2$H emission appears to be robust.

\subsubsection{DM Tau}
\label{sec:dmt}

For DM~Tau, the comparison of our best fit model emission predictions to observations is shown in Fig.~\ref{fig:best}.
 As shown in Fig.~\ref{fig:hires} and discussed in \S~3, DM Tau has two C$_2$H rings.   The inner ring is at R $\sim 50$~AU and the outer weaker ring is observed at $R \sim 350$~AU.  
Fig.~\ref{fig:dmt} shows distribution of physical and chemical parameters from our model that best matches the DM Tau emission distribution, which is compared to observations in Fig.~\ref{fig:best}.   As in TW Hya the outer C$_2$H ring appears at the edge of the pebble disk traced by 1.3 mm-emission as there 
is fainter ($3\sigma$) extended emission seen in our 1.3 mm image out to 300~AU.   It is this point where the UV-dominated disk can begin to penetrate to greater depth and enhance the column of UV-exposed gas and dust.   Our models have some difficulty matching the sharpness of the outer disk ring, though the observational data is limited by the noise level in the outer disk.   

 As noted in \S~3.3 the inner C$_2$H emission ring observed in DM~Tau could be due to either dust optical depth or perhaps the beginning of carbon depletion in the inner disk such as seen in TW Hya.  Due to this uncertainty we did not model this central hole in greater detail.  However, the inner disk still has strong C$_2$H emission, which  is clearly distinct from TW Hya.   An additional notable difference is that DM Tau has a larger abundance of elemental carbon present in the gaseous state.   Based on HD measurements TW Hya and DM Tau have roughly comparable gas masses \citep{McClure16}.  However, the CO mass derived for DM Tau from C$^{18}$O observations is a factor of 18 higher than TW Hya \citep{jWilliams14}.  Without significant carbon depletion, our models (e.g. \reffig{figCMP}) predict strong central emission consistent with observations.   
 Thus, we interpret the strength of the central emission in DM Tau as a likely time-dependent effect.   The combined effects of settling and radial drift will have a stronger influence on the outer disk.  The lower density outer disk will also be more transparent to UV radiation than the denser inner disk.  Hence the effects of carbon extraction from CO would have greater efficiency in the outer disk.  Over time these effects will become more significant in the inner disk until the system evolves toward the central carbon depletion as seen in TW Hya.

 \subsection{Excitation Analysis}

From our modeling of TW Hya (\reffig{fig:best} and \reffig{fig:twh}) and the contribution function (\reffig{fig:contrib}) we predict that C$_2$H is emitting from layers with n$_{\rm H_2}$ $\sim 10^7 - 10^8$ cm$^{-3}$ in the warm (T$_{gas}$ $\sim 40$~K) surface layers, with  additional contribution from a denser midplane.  
\citet{Kastner15} used an analysis of the N = 4 -- 3/N = 3 -- 2 ratio to constrain the excitation conditions of the C$_2$H emission.   Based on their analysis the emission originates from relatively warm (T $>$ 40~K) but lower density, $n_{\rm H_2} \ll 10^7$ cm$^{-3}$ gas.  In this instance, for the higher transition data, an unresolved line detection from APEX was used, while the N = 3 -- 2 emission was resolved by the SMA with a resulting line ratio estimation of 0.73 (plus 10\% errors).   
Our higher resolution data finds a much greater line flux in the N = 4 -- 3 transition by a factor of 3  (see disk integrated flux in Table~\ref{tab:obs}), which is  different from that estimated by \citet{Kastner15}, presumably due to beam dilution. 

Thus the flux density ratio is increased, with a value of:

\begin{equation}
\rm
\frac{N = 4 -- 3, J = \frac{9}{2} - \frac{7}{2}, F = 4-3}{N = 3 -- 2, J = \frac{9}{2} - \frac{7}{2}, F = 4-3} = 2.16 \pm 0.45.\footnote{We have assumed 15\% calibration uncertainty for both ALMA and SMA.}   
\end{equation}

\begin{figure}[b]
\begin{centering}
\includegraphics[width=0.5\textwidth]{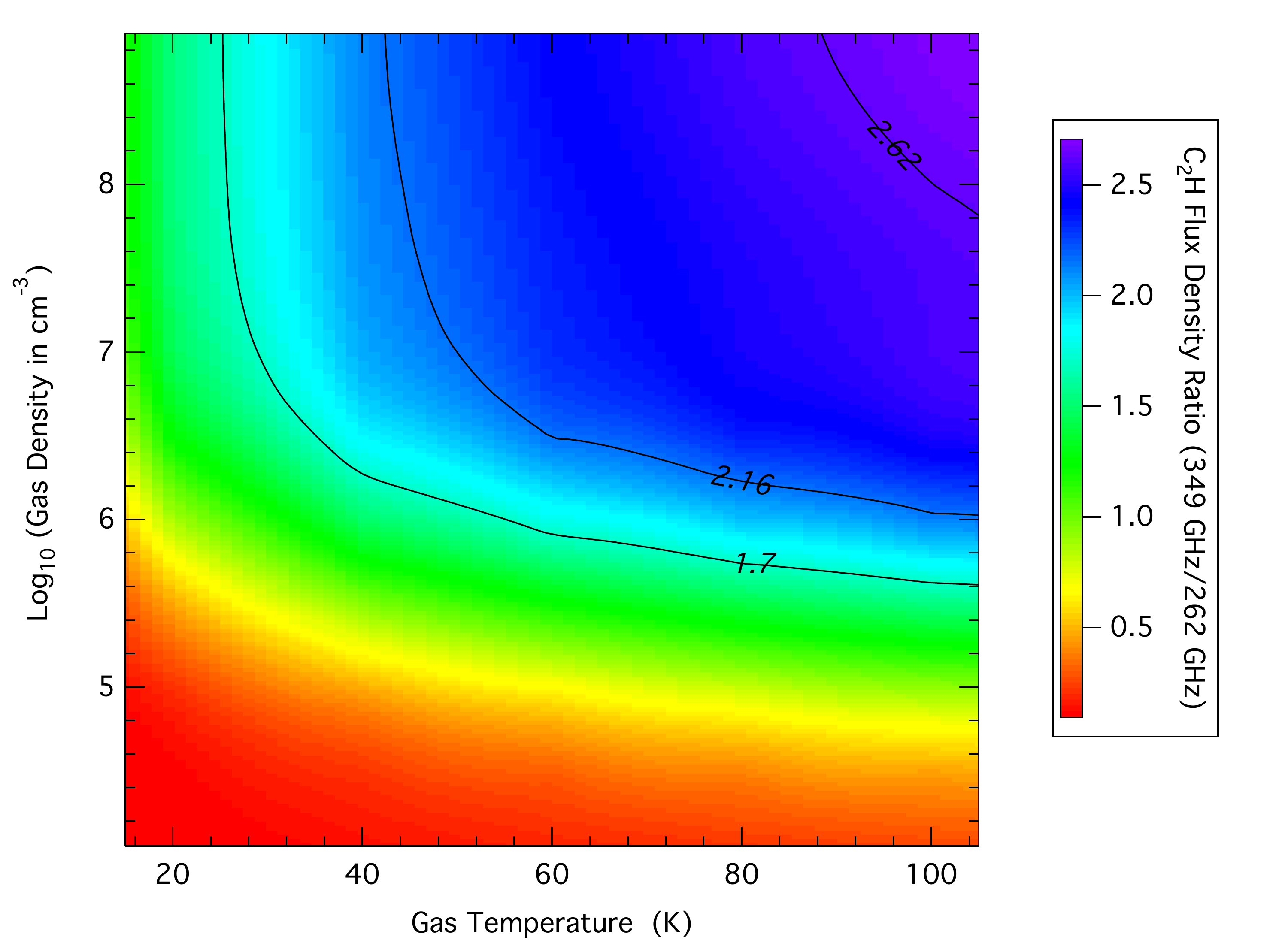} 
\caption{Excitation calculations for the flux density ratio of the
N = 4 -- 3, J = $\frac{9}{2} - \frac{7}{2}$, F = 4-3 to N = 3 -- 2, J = $\frac{9}{2} - \frac{7}{2}$, F = 4-3 transitions.  The observed ratio (2.16) with 1$\sigma$ error bars is shown as contours \label{fig:xcit}}
\end{centering}
\end{figure}

\noindent We have  therefore re-analyzed the excitation characteristics of this emission using RADEX \citep{radex} and the state to state collision rates by \citet{c2hcolxs}. This analysis is shown in Fig.~\ref{fig:xcit}.  With this revision, it seems clear that the emission arises from gas with at least moderate density $>$ 10$^6$ cm$^{-3}$.  If applicable to C$_2$H, this is consistent with fairly high densities.  At face value, the observed ratio is also better fit by emission within warmer gas ($\sim 40 - 50$~K).   

We have also explored solutions including a dust rich background source, as the C$_2$H emission lies in front of the central dust rich midplane emitting at a temperature below 20 K at these radii \citep{Andrews16}.  This compresses the dependences at low densities (for all temperatures) and low temperatures (for all densities) but does not change the conclusion: C$_2$H appears to be emitting from gas with densities greater than 10$^6$~cm$^{-3}$ consistent with our predictions.   From \reffig{fig:dmt} we would predict that the outer C$_2$H ring in DM Tau is emitting from gas that has lower density ($3 \times 10^4$~cm$^{-3} < {\rm n_{H_2}} < 10^6$~cm$^{-3}$) and is colder (T$_{gas} \sim 10-20$~K).

\subsection{Photodissociation of Carbonaceous Grains as the Origin of Simple Hydrocarbons}
\label{sec:pah}

In this paper we have outlined a consistent picture of coupled dust + gas physical chemical evolution that leads to the creation
of hydrocarbon rings.   We have implicitly assumed that C$_2$H
is produced from carbon that was extracted from CO, aided by the depletion of oxygen in water ice.
On the other hand, as discussed by 
\citet{Kastner15} the photo-ablation of aliphatic or aromatic hydrocarbon grains would be a ready
source of carbon as carbon-rich grains contain nearly half of the cosmic abundance of carbon 
\citep{Jones13, Chiar13}.    Moreover, if carbon grains are being destroyed in gas where water remains frozen, then the C/O ratio can be $>$ 1.   

\begin{figure*}[htbp]
\centering
\includegraphics[width=1.0\linewidth]{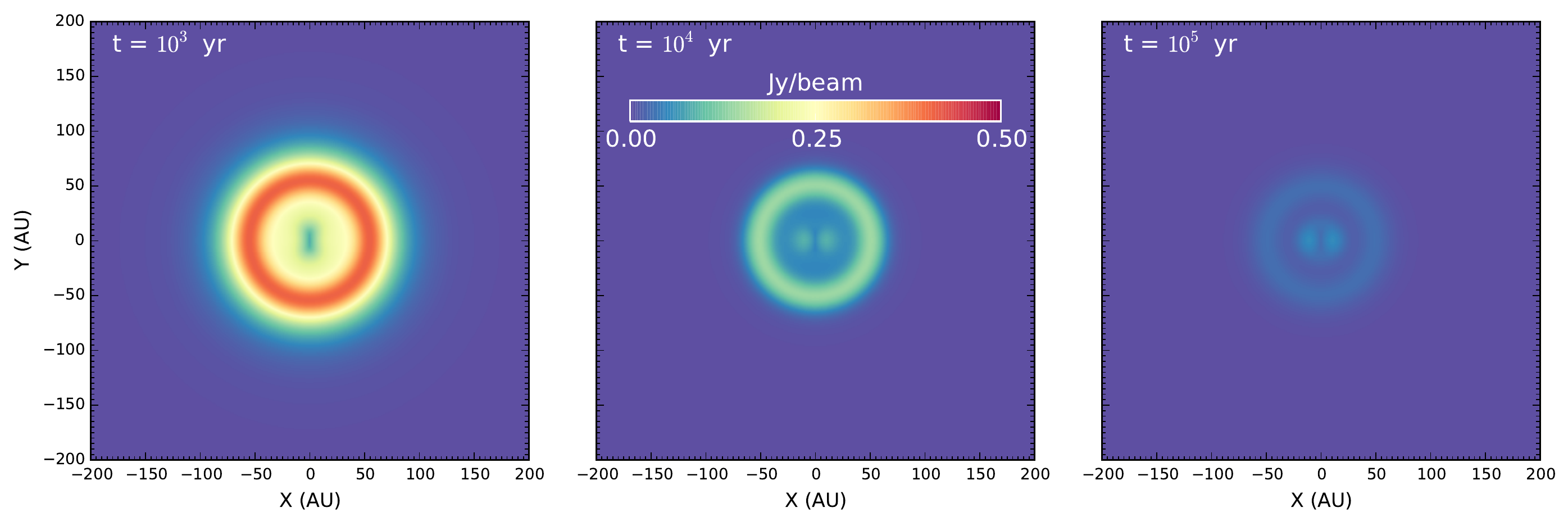}
\caption{The \ce{C2H} $J{=}7/2{-}5/2$, $F{=}4{-}3$ \& $3{-}2$ emission at $10^3$, $10^4$, and $10^5$ years, starting
with a uniformly high \ce{C2H} abundance of $10^{-5}$.  The underlying disk
structure is the one specified in \reftab{tab:model} for TW~Hya.}
\label{fig:pah}
\end{figure*}

To explore this issue we use a calculation of the equilibrium abundance of C$_2$H if produced via photodissociation of hydrogenated amorphous carbon grains, one of the potential carriers of refractory carbon, and destroyed by atomic oxygen.  There are no calculations of  photodissociation channels of hydrogenated amorphous carbon grains (a-C:H) that directly lead to C$_2$H.  However, laboratory work and models of hydrogenated amorphous carbon photoablation and carbon chemistry by \citet{Alata15} show that creation of other hydrocarbons (CH$_4$, C$_2$H$_2$ as examples) leads to an increase in the abundance of C$_2$H.  For this purpose we adopt their total photodissociation rate regardless of the product.  This is $k_{ph} = 2 \times 10^{-12} G_0 \exp(-2.0A_V) F_i$~s$^{-1}$.
 For TW Hya,  $G_0 is \sim 320$ at 100 AU \citep{bergin_lyalp}, including the contribution from Ly $\alpha$ this would increase to $\sim 2000$ \citep{herczeg_twhya1, bergin_h2}.  $F_i$ is a factor that accounts for the fact that the parent a-C:H grain being ablated each time a photon is capable of breaking off a fragment. For simplicity we adopt $F_i = 0.5$.   The unshielded photorate is consistent with the calculation of polycyclic aromatic hydrocarbon (PAH) photodissociation producing C$_2$, C$_2$H, or C$_2$H$_2$ fragments for a PAH consisting of $<$ 20 atoms \citep{Visser07}.  Thus, this calculation is roughly appropriate for PAHs, which are another potential solid state carbon carrier. 
     
   To obtain the maximum amount of C$_2$H that might be created with this rate as the source term we further assume that C$_2$H is the ultimate product of all dissociations.  Under this case the equilibrium abundance will be as follows:

\begin{equation}
\frac{dn_{\rm C_2H}}{dt} = n_{\rm a-C:H}k_{ph}  - n_{\rm C_2H} n_{\rm O} k_{\rm O, C_2H} 
\end{equation}

\noindent Here $k_{\rm O, C_2H} = 10^{-10}$ cm$^3$/s is the destruction rate of C$_2$H with oxygen atoms (assuming this is the main destruction channel).
In steady state both right hand terms are equivalent and we can solve for the C$_2$H abundance:

\begin{equation}
x_{\rm C_2H} = \frac{x_{\rm a-C:H} k_{ph}}{x_{\rm O}n_{\rm H_2} k_{\rm O, C_2H} }
\end{equation}

\noindent We assume that the abundance of carbon grains is $\sim 3 \times 10^{-7}$ 
relative to H and each grain carries $\sim 50$ carbon atoms as determined for PAHs \citep{tielens_book}.
Strictly PAHs are not the same chemical form as a-C:H, but it provides a good baseline for the amount of carbon that might be present in the the solid state.  We also assume that the distance is 60 AU, $A_V = 0.5^m$ (or $\tau_{UV} = 1$), and a gas density of 10$^8$~cm$^{-3}$ which is the density and UV optical depth seen in the C$_2$H layers in our models.    Using these expressions if the abundance of oxygen atoms is near cosmic ($10^{-4}$) then the equilibrium abundance of C$_2$H would be $\sim 5 \times 10^{-10}$.  Based on our simulations, this would be well below the value of $\sim 10^{-7}$ needed for detection.  To be consistent with our results (with the H$_2$ gas mass constrained by HD), the oxygen abundance would need to be reduced by 2-3 orders of magnitude.  Furthermore, species such as PAHs are difficult to detect in T Tauri systems with the suggestion their abundances are reduced by 10-100 orders of magnitude \citep{geers07, Akimkin13}; this would require even lower abundances of atomic oxygen.   Thus, if carbonaceous grains are the source term for C$_2$H emission, it also requires reduced abundances of oxygen in the disk atmosphere.

The timescale to reduce the grain by one small fragment is short and is only $\sim 100$ years.  Depending on the size, small grains  could be destroyed on timescales below 10$^5$ yrs \citep{Alata15}, but large grains or smaller grains in deeper layers would survive.  However, the  photorates of large  ($>$ 25 atoms) grains are significantly reduced by many orders of magnitude \citep{Visser07}  as there are more internal models to share the energy, as opposed to breaking a bond.  Thus very large grains may not be the most viable source terms for C$_2$H production. 

  If a-C:H or PAHs were the source term another question would be how long would C$_2$H last in the gaseous state.    Since our chemical network does
not explicitly include PAH chemistry, we mimic the effect of PAH/a-C:H
photodissociation by artificially setting the initial abundance of \ce{C2H}
to $10^{-5}$ (relative to hydrogen nuclei).  This is consistent with the carbon grain abundance adopted above, assuming 50 carbon atoms which then uniformly produce C$_2$H.  Otherwise this model assumes a normal C/O ratio.  We then evolve the system for 1~Myr to see how the resulting emission in \ce{C2H} changes with time.
 
\reffig{fig:pah} shows the \ce{C2H} emission map at $t=10^3$, $10^4$, and $10^5$
years.  Here the \ce{C2H} emission decreases with time monotonically and only approaches
observed values at  early stages ($\sim 10^{3}$~yrs).  As in the discussion above longer timescales would be found for lower oxygen abundances.   However, without a continuous supply of small PAH/a-C:H grains as a source of hydrocarbons and without depletion of oxygen, the carbon budget contained in the initial hydrocarbons (such as \ce{C2H}) will quickly be converted into other species (such as CO through oxygen production via H$_2$O ice photodesorption).
As noted in \citet{Kastner15}, the estimated PAH abundance is barely enough to account for the observed \ce{C2H} emission, especially if we consider that the PAH abundance is usually argued to be already depleted in the disk environment \citep{Akimkin13}.   Thus, while destruction of carbon grains may be a source of carbon for C$_2$H, it is likely that the production of C$_2$H is fueled by the extraction of carbon from CO and depletion of oxygen from CO and other major oxygen-bearing species (such as water and \ce{CO2}).  This is consistent with the measured depletion of CO in TW Hya at the very least \citep[][]{favre13a, Nomura16, Schwarz16}; we therefore predict that the outer disk of DM Tau should also exhibit a reduction in the abundance of volatile CO, while the inner disk must have high carbon content to remain consistent with C$^{18}$O measurements \citep{jWilliams14}

\subsection{A physical or a chemical effect?}

In our model we have associated dust evolution to two effects.  One is physical, that is a redistribution of the UV opacity that increases the penetration of UV photons in the disk system.  The other is the relative trapping of volatiles as icy mantles coating grains that settle, grow, and drift which induces an increase in the C/O ratio.  This brings up the question as to whether one can reproduce these observations with an increased C/O ratio alone. The work by \citet{Kama16a} provides some insight to this issue as they explored the emission of several carbon-bearing species, including an unresolved observation of C$_2$H, in TW Hya.  Similar to our result they find that reproducing the emission requires C/O $> 1$.  Their model does include a vertically stratified dust model that accounts for the settling of larger grains, but does not include the effects of grain drift.  At face value \citet{Kama16a} show that the total C$_2$H flux can be reproduced by increasing the C/O ratio.  

In our models for TW Hya alone we require central carbon depletion to reproduce the emission distribution.  Thus it would be hard to directly point out the effects of the UV field in this case.  However, we have the two rings seen in DM Tau.  The outer ring in particular requires both UV to power the chemistry and C/O $> 1$.  In addition, our models, for both sources, and \citet{Kama16a} have a stronger UV field on the disk surface due to dust settling; in our case we find $G_0 \ge 1$ in layers where C$_2$H is forming.  
  This, and the known association of C$_2$H (and other hydrocarbons) with photodissociation regions \citep{Pilleri13, Nagy15}, argues that the UV field is also an important factor in hydrocarbon production.

\section{Summary and Implications}\label{summary}

We have presented high-resolution observations of C$_2$H and \ce{C3H2} in the protoplanetary disks of TW Hya and DM Tau, two classical T Tauri stars.  We summarize here key observational facts and results from a detailed analysis:

\begin{enumerate}

\item  In TW Hya we resolve a single emission ring that peaks near 55 AU in the sub-millimeter emission lines of both C$_2$H and $c$-C$_3$H$_2$.  For C$_2$H the central emission hole is resolved with emission significantly weaker (by factor of 6) in the disk center than at ring peak. The peak emission in the hydrocarbon ring sits just outside of the resolved sub-mm dust disk.

\item  The central hole of \ce{C2H} in \TWH{} requires depletion of carbon within $\sim$40 AU, which is consistent with previous observations of CO isotopologues. We further rule out high dust optical depth as the cause of the ring structure, and require carbon depletion as the cause.   

\item In DM Tau we detect two rings: an inner ring that is centered on a bright central peak in the sub-mm continuum emission (but not an edge) near $R\sim50$ AU and an outer ring at 350~AU near the edge of the detectable disk in sub-mm continuum emission.

\item We suggest that hydrocarbon rings are created via the coupling of the gas-phase chemistry to the spatial evolution of ice-coated dust grains.   As ice-coated dust grains settle to the dust rich midplane they grow to larger sizes whereupon they drift towards smaller radii.   This concentrates the dust mass in the midplane within the inner tens of AU of the disk, with two attendant effects that are ingredients for the production of hydrocarbon rings.  First, the redistribution of dust mass also redistributes the UV opacity, leaving the disk surface and outer disk essentially a UV-photon-dominated PDR.   Second, it also concentrates the volatiles, with a greater effect for the more tightly bound H$_2$O than the more volatile CO, enhancing the gas-phase C/O ratio to values $> 1$ and increasing the hydrocarbon emission.   This should readily deplete the outer disk of volatiles, which would therefore be the first ``observable'' effect.    In the inner disk carbon could also be depleted due to formation of refractory grains or sequestration in larger bodies that are less susceptible to evaporation, which creates an inner hydrocarbon hole.

\item Using ancillary data, we perform an excitation analysis on the C$_2$H emission showing that the emission must arise from gas with n$_{\rm H_2} > 10^6$~cm$^{-3}$.  This is consistent with our model predictions.

\item We present parametric models that match the observed emission distribution in both systems to within a factor of 1.5.  They illustrate the importance of chemical effects for hydrocarbon ring formation. 
\end{enumerate}

The presence of a hydrocarbon ring in two disk systems of disparate sizes and ages does not, by itself, provide evidence that hydrocarbon rings are a widespread feature of disk chemical evolution.  Given the sensitivity of ALMA this question will be answered with a larger statistically significant sample.
However, we suggest that these effects are likely prevalent.
First, \citet{Kastner14} performed an unbiased line survey between 275 - 357 GHz of TW Hya and V4046 Sgr finding that C$_2$H lines are among the strongest in the band (even brighter than $^{13}$CO).  This hints at a significant \ce{C2H} abundance enhancement.  Second, given the coincidence of the C$_2$H rings with the edge of the pebble disk, the production of hydrocarbon rings  is most likely tied to dust settling, growth, and migration.   These are common physical effects that have strong observational support in a wide range of systems \citep{gd98, furlan06, pdg07, isella07, andrews12}.
            Thus we predict that hydrocarbon rings will be a common feature in disks, and the coupled gas-dust evolution is the single most important factor that influences the observable chemical signatures of gas-rich disk systems.  

\acknowledgements

This work was supported by funding from the National Science Foundation grant AST-1514670 and AST-1344133 (INSPIRE) along with NASA XRP grant NNX16AB48G.  This paper makes use of the following ALMA data:
  ADS/JAO.ALMA\#2013.1.00198.S. ALMA is a partnership of ESO (representing
  its member states), NSF (USA) and NINS (Japan), together with NRC
  (Canada) and NSC and ASIAA (Taiwan) and KASI (Republic of Korea), in 
  cooperation with the Republic of Chile. The Joint ALMA Observatory is 
  operated by ESO, AUI/NRAO and NAOJ.  The National Radio Astronomy Observatory is a facility of the National Science Foundation operated under cooperative agreement by Associated Universities, Inc.
  We are grateful to the anonymous referee for numerous thoughtful and useful comments.

\appendix
\renewcommand\thefigure{\thesection.\arabic{figure}}    
\setcounter{figure}{0}
\section{Empirical Approximation for Elemental Abundance Distribution}

To model the depletion of elements, we assume that the scale height for the abundance of the element under consideration is a fraction of the scale height
of hydrogen at a given radius.  Specifically, if the density of the hydrogen gas as a function of vertical height $z$ and radius $r$ is $n(z,r)$, then the abundance of the element to be depleted relative to its canonical value is taken to be
\begin{equation} X(z,r) = \left[\frac{n(z,r)}
{n(z=0,r)}\right]^{f_\text{depl}(r)}.
\label{eqdep1}
\end{equation}
For $f_\text{depl}$ we use the following form
\begin{equation}
  f_\text{depl}(r) = \frac{1}{y^2} - 1,
\label{eqdep2}
\end{equation} 
where
\begin{equation}
  y(r) = b + \frac{a-b}{2} \left[1 - \tanh\left(\frac{r-r_0}{r_\text{s}}\right)\right].
\label{eqdep3}
\end{equation} 
Here $a$, $b$, $r_0$, and $r_\text{s}$ are parameters, with $a$ and $b$ being dimensionless.  $a = y_\text{max} =
y|_{r\to-\infty}$, and $b = y_\text{min} = y|_{r\to+\infty}$.  $r_0$ marks the
location outside of which the element is depleted, and within which may be
enhanced, and $r_\text{s}$ quantifies the width of the transition region.  If
$y>1$ (which may occur when $a>1$), $f_\text{depl}$ can take negative values,
which means this parameterization allows the elemental distribution in the
inner disk to increase with height above the disk.  $f_\text{depl} = 0$ means
no depletion, and the larger $f_\text{depl}$, the more the element is depleted.
To see the physical meaning of $f_\text{depl}$, assume $$n(z,r)\propto
\exp\left[-\frac{z^2}{2h^2(r)}\right],$$ where $h(r)$ is the scale height.
Then we have
$$X(z,r) \propto
\exp\left\{-\frac{z^2}{2\left[h(r)/f^{1/2}_\text{depl}(r)\right]^2}\right\}.$$
So $f^{1/2}_\text{depl}$ is the ratio between the scale height of hydrogen gas and the scale heights of the abundances of oxygen or carbon.  The calculated
elemental abundances for carbon and oxygen are used as input to the thermo-chemical solver at each location of the disk.  The parameters are chosen heuristically for each disk to investigate what C and O distribution best reproduce the observed C$_2$H observations.

\begin{figure*}[htbp]
\centering
\includegraphics[width=1.0\linewidth]{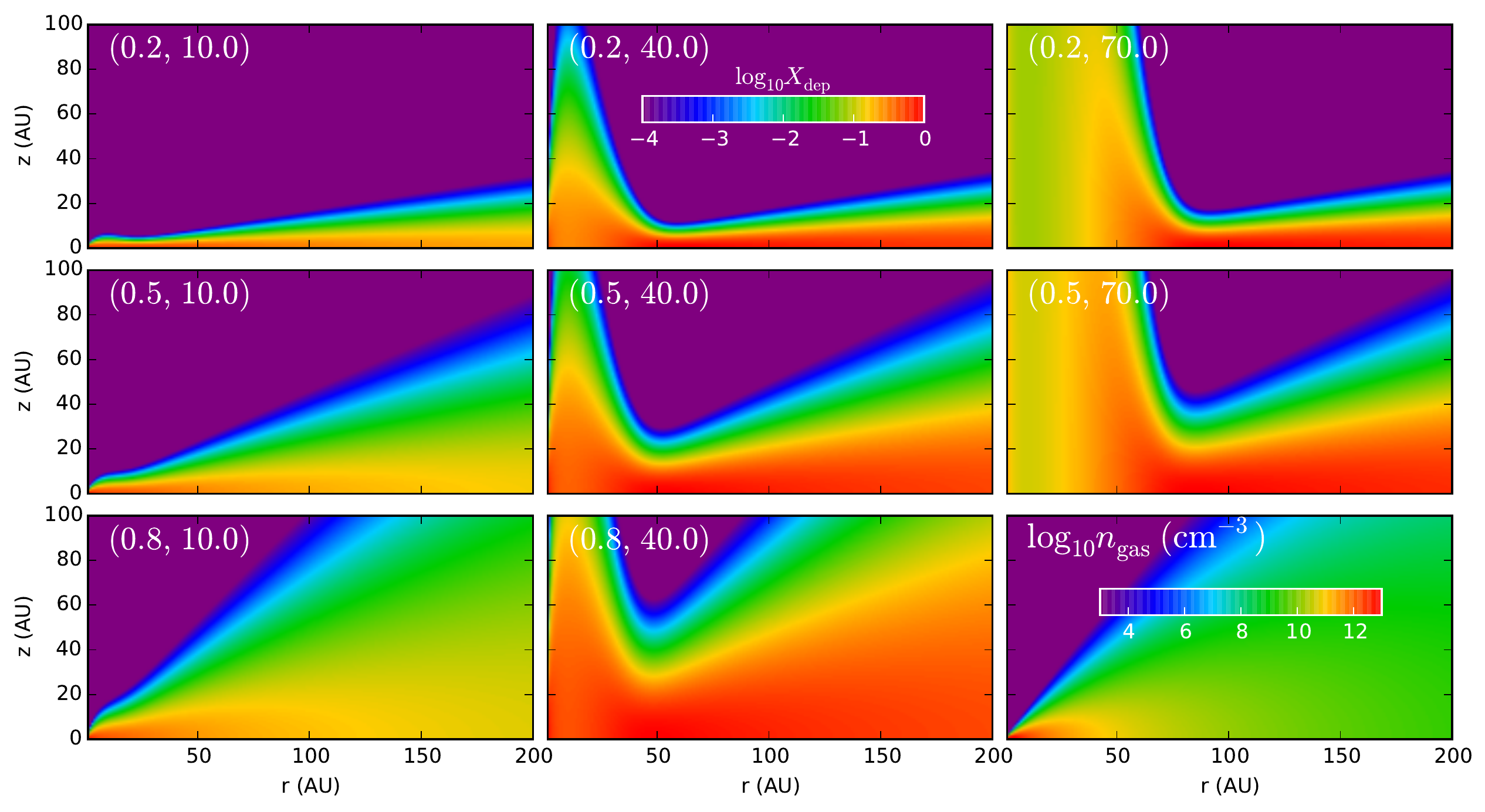}
\caption{Panels illustrate  depletion profiles for different values of (b, $r_0$) ($r_0$ is in AU), while (a, $r_s$) are fixed to (1.0, 10.0 AU) (which are close to the values used in the models).  These parameters are defined in equations (A1--A3).  The lower-right panel shows the adopted density distribution used for this illustration (which is needed because in our empirical model the degree of depletion depends on the scale height of the gas distribution, as quantitatively described by equation~A1).  The density structure is calculated with the following parameters: $r_c$ = 50.0 AU, $\gamma$ = 1.5, $h_c$ = 10.0 AU, and $\psi = 0$ \citep[using the parameterization given by][]{Andrews09}.  The color scale of all panels are in log scale.  For the elemental distribution, the highest value (in red) is to be interpreted as the non-depleted ISM abundance.
}
\label{fig:eled}
\end{figure*}

In Fig.~\ref{fig:eled} we provide example distributions (in a relative sense) as to how this parameterization operates.  Different panels correspond to different values of the $b$ and $r_0$ parameters, while $a$ is fixed to 1 and $r_\text{s}$ fixed to 10~AU.  The lower-right panel is the gas density distribution used for the illustrations, which is needed per equation~(A1).  Basically, the smaller the $b$ parameter is, the more the element is vertically depleted; and with a greater value for the $r_0$ parameter, the depletion will starts to occur at a greater distance, which means that $r_0$ is a kind of a ``snow line'' of the major carriers of the element under consideration.

\newpage

\setcounter{figure}{0}

\section{C$_2$H Contribution Function in TW Hya}

We define the contribution of each location to the total emission to be the
power emitted per unit length per unit area along the line of sight at that
location that is actually received by a distant observer.  From this intuitive
definition, we see that the contribution function depends on the line of sight,
and is a 3-D function in general.  Even if a location is very emissive, if it
is behind high optical depth along the line of sight, its contribution will be
low.  Mathematically, the contribution function we use is
\begin{equation}
  \text{contrib}(\mathbf{x}) = j(\mathbf{x}) e^{-\tau_\text{los}(\mathbf{x})},
\end{equation} 
where \(j(\mathbf{x})\) is the emission coefficient at \(\mathbf{x}\), while
\(\tau_\text{los}(\mathbf{x})\) is the optical depth measured from
\(\mathbf{x}\) to infinity along the line of sight.

For a disk with axisymmetry, the contribution function will not be axisymmetric
unless the disk is viewed face on, in which case we can integrate it vertically
(and times \(r\)) to get the radial contribution function, and then integrate
radially to get the radial accumulative contribution.

Usually we only care about the relative importance of each location to the
total observed emission, so we normalize the contribution function by dividing
it with its maximum value.

\begin{figure*}[htbp]
\centering
\includegraphics[width=0.6\linewidth]{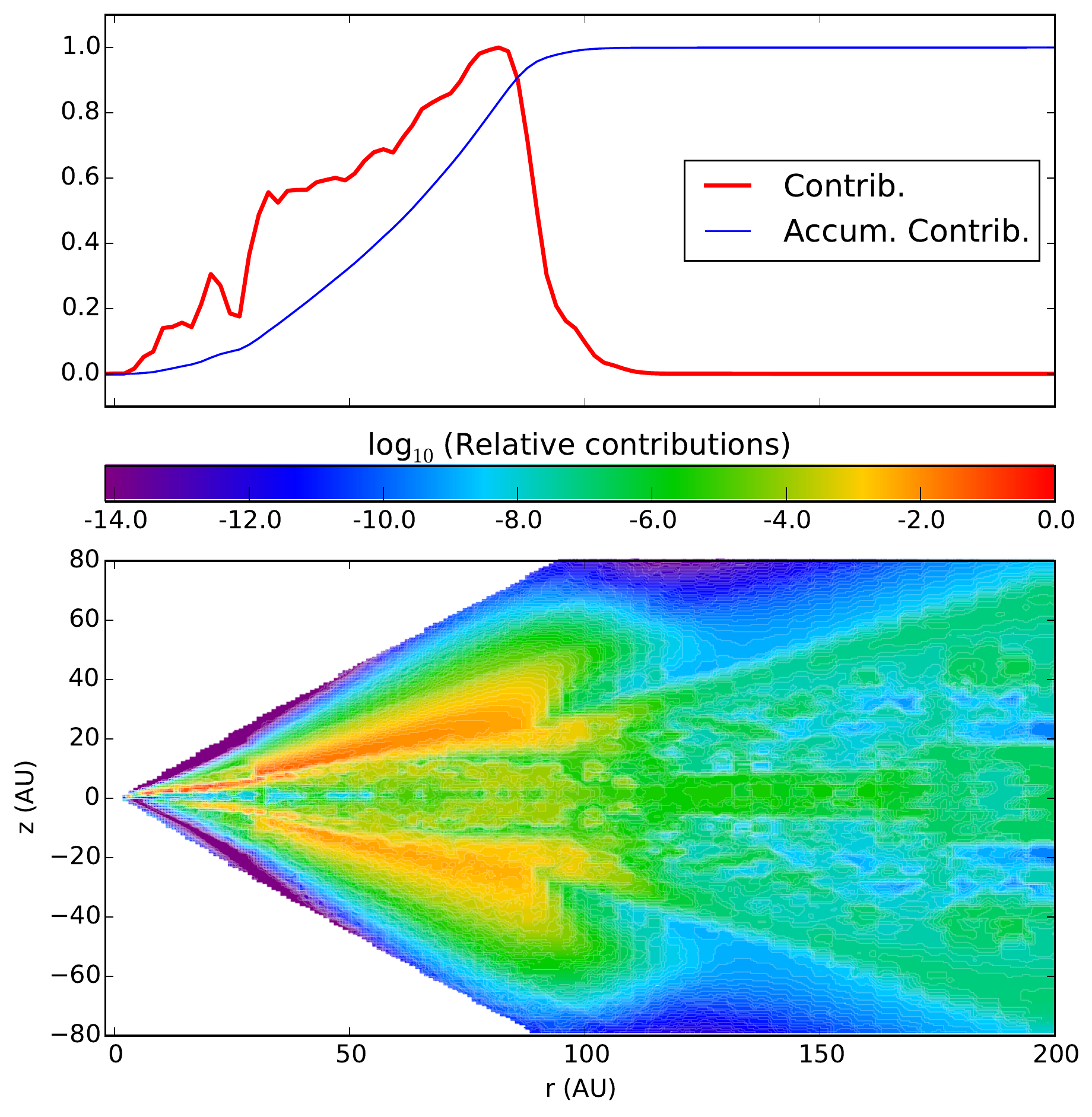}
\caption{{\em Top Panel:} 
Contribution function at each radius (vertically integrated), and the radial accumulated contribution
from the TW Hya model shown in Figs.~\ref{fig:best} and \ref{fig:twh}.
{\em Bottom panel: Relative contribution to the C$_2$H line emission as a function of
position in the disk.}.  Definitions described in Appendix.  
}
\label{fig:contrib}
\end{figure*}

\bibliographystyle{apj}
\bibliography{ted}

\begin{thebibliography}{}
\expandafter\ifx\csname natexlab\endcsname\relax\def\natexlab#1{#1}\fi

\bibitem[{{Aikawa} \& {Nomura}(2006)}]{an06}
{Aikawa}, Y., \& {Nomura}, H. 2006, \apj, 642, 1152

\bibitem[{{Aikawa} {et~al.}(1999){Aikawa}, {Umebayashi}, {Nakano}, \&
  {Miyama}}]{aikawa99}
{Aikawa}, Y., {Umebayashi}, T., {Nakano}, T., \& {Miyama}, S.~M. 1999, \apj,
  519, 705

\bibitem[{{Aikawa} {et~al.}(2002){Aikawa}, {van Zadelhoff}, {van Dishoeck}, \&
  {Herbst}}]{aikawa_vanz02}
{Aikawa}, Y., {van Zadelhoff}, G.~J., {van Dishoeck}, E.~F., \& {Herbst}, E.
  2002, \aap, 386, 622

\bibitem[{{Akimkin} {et~al.}(2013){Akimkin}, {Zhukovska}, {Wiebe}, {Semenov},
  {Pavlyuchenkov}, {Vasyunin}, {Birnstiel}, \& {Henning}}]{Akimkin13}
{Akimkin}, V., {Zhukovska}, S., {Wiebe}, D., {et~al.} 2013, \apj, 766, 8

\bibitem[{{Alata} {et~al.}(2015){Alata}, {Jallat}, {Gavilan}, {Chabot},
  {Cruz-Diaz}, {Munoz Caro}, {B{\'e}roff}, \& {Dartois}}]{Alata15}
{Alata}, I., {Jallat}, A., {Gavilan}, L., {et~al.} 2015, \aap, 584, A123

\bibitem[{{Andrews}(2015)}]{Andrews15}
{Andrews}, S.~M. 2015, \pasp, 127, 961

\bibitem[{{Andrews} {et~al.}(2011{\natexlab{a}}){Andrews}, {Wilner},
  {Espaillat}, {Hughes}, {Dullemond}, {McClure}, {Qi}, \& {Brown}}]{Andrews11}
{Andrews}, S.~M., {Wilner}, D.~J., {Espaillat}, C., {et~al.}
  2011{\natexlab{a}}, \apj, 732, 42

\bibitem[{{Andrews} {et~al.}(2011{\natexlab{b}}){Andrews}, {Wilner},
  {Espaillat}, {Hughes}, {Dullemond}, {McClure}, {Qi}, \&
  {Brown}}]{Andrews2011}
---. 2011{\natexlab{b}}, \apj, 732, 42

\bibitem[{{Andrews} {et~al.}(2009{\natexlab{a}}){Andrews}, {Wilner}, {Hughes},
  {Qi}, \& {Dullemond}}]{Andrews2009}
{Andrews}, S.~M., {Wilner}, D.~J., {Hughes}, A.~M., {Qi}, C., \& {Dullemond},
  C.~P. 2009{\natexlab{a}}, \apj, 700, 1502

\bibitem[{{Andrews} {et~al.}(2009{\natexlab{b}}){Andrews}, {Wilner}, {Hughes},
  {Qi}, \& {Dullemond}}]{Andrews09}
---. 2009{\natexlab{b}}, \apj, 700, 1502

\bibitem[{{Andrews} {et~al.}(2012){Andrews}, {Wilner}, {Hughes}, {Qi},
  {Rosenfeld}, {{\"O}berg}, {Birnstiel}, {Espaillat}, {Cieza}, {Williams},
  {Lin}, \& {Ho}}]{andrews12}
{Andrews}, S.~M., {Wilner}, D.~J., {Hughes}, A.~M., {et~al.} 2012, \apj, 744,
  162

\bibitem[{{Andrews} {et~al.}(2016){Andrews}, {Wilner}, {Zhu}, {Birnstiel},
  {Carpenter}, {P\'{e}rez}, {Bai}, {\"{O}berg}, {Hughes}, {Isella}, \&
  {Ricci}}]{Andrews16}
{Andrews}, S.~M., {Wilner}, D.~J., {Zhu}, Z., {et~al.} 2016, \apj, in press.

\bibitem[{{Bergin} {et~al.}(2003){Bergin}, {Calvet}, {D'Alessio}, \&
  {Herczeg}}]{bergin_lyalp}
{Bergin}, E., {Calvet}, N., {D'Alessio}, P., \& {Herczeg}, G.~J. 2003, \apjl,
  591, L159

\bibitem[{{Bergin} {et~al.}(2004){Bergin}, {Calvet}, {Sitko}, {Abgrall},
  {D'Alessio}, {Herczeg}, {Roueff}, {Qi}, {Lynch}, {Russell}, {Brafford}, \&
  {Perry}}]{bergin_h2}
{Bergin}, E., {Calvet}, N., {Sitko}, M.~L., {et~al.} 2004, \apjl, 614, L133

\bibitem[{{Bergin} {et~al.}(2014){Bergin}, {Cleeves}, {Crockett}, \&
  {Blake}}]{Bergin14}
{Bergin}, E.~A., {Cleeves}, L.~I., {Crockett}, N., \& {Blake}, G.~A. 2014,
  Faraday Discuss., 168, 61

\bibitem[{{Bergin} {et~al.}(2010){Bergin}, {Hogerheijde}, {Brinch},
  {et~al.}}]{bergin10b}
{Bergin}, E.~A., {Hogerheijde}, M.~R., {Brinch}, C., {et~al.} 2010, \aap, 521,
  L33+

\bibitem[{{Bergin} {et~al.}(2013){Bergin}, {Cleeves}, {Gorti}, {Zhang},
  {Blake}, {Green}, {Andrews}, {Evans}, {Henning}, {{\"O}berg}, {Pontoppidan},
  {Qi}, {Salyk}, \& {van Dishoeck}}]{bergin_hd}
{Bergin}, E.~A., {Cleeves}, L.~I., {Gorti}, U., {et~al.} 2013, \nat, 493, 644

\bibitem[{{Bethell} \& {Bergin}(2011)}]{bb11a}
{Bethell}, T.~J., \& {Bergin}, E.~A. 2011, \apj, 739, 78

\bibitem[{{Birnstiel} {et~al.}(2012){Birnstiel}, {Klahr}, \&
  {Ercolano}}]{Birnstiel12}
{Birnstiel}, T., {Klahr}, H., \& {Ercolano}, B. 2012, \aap, 539, A148

\bibitem[{{Blevins} {et~al.}(2016){Blevins}, {Pontoppidan}, {Banzatti},
  {Zhang}, {Najita}, {Carr}, {Salyk}, \& {Blake}}]{Blevins16}
{Blevins}, S.~M., {Pontoppidan}, K.~M., {Banzatti}, A., {et~al.} 2016, \apj, in
  press.

\bibitem[{{Calvet} {et~al.}(2005){Calvet}, {D'Alessio}, {Watson},
  {Franco-Hern{\'a}ndez}, {Furlan}, {Green}, {Sutter}, {Forrest}, {Hartmann},
  {Uchida}, {Keller}, {Sargent}, {Najita}, {Herter}, {Barry}, \&
  {Hall}}]{Calvet05}
{Calvet}, N., {D'Alessio}, P., {Watson}, D.~M., {et~al.} 2005, \apjl, 630, L185

\bibitem[{{Chapillon} {et~al.}(2008){Chapillon}, {Guilloteau}, {Dutrey}, \&
  {Pi{\'e}tu}}]{Chapillon08}
{Chapillon}, E., {Guilloteau}, S., {Dutrey}, A., \& {Pi{\'e}tu}, V. 2008, \aap,
  488, 565

\bibitem[{{Chiang} \& {Youdin}(2010)}]{Chiang10}
{Chiang}, E., \& {Youdin}, A.~N. 2010, Annual Review of Earth and Planetary
  Sciences, 38, 493

\bibitem[{{Chiang} {et~al.}(2001){Chiang}, {Joung}, {Creech-Eakman}, {Qi},
  {Kessler}, {Blake}, \& {van Dishoeck}}]{Chiang01}
{Chiang}, E.~I., {Joung}, M.~K., {Creech-Eakman}, M.~J., {et~al.} 2001, \apj,
  547, 1077

\bibitem[{{Chiar} {et~al.}(2013){Chiar}, {Tielens}, {Adamson}, \&
  {Ricca}}]{Chiar13}
{Chiar}, J.~E., {Tielens}, A.~G.~G.~M., {Adamson}, A.~J., \& {Ricca}, A. 2013,
  \apj, 770, 78

\bibitem[{{Cleeves} {et~al.}(2013){Cleeves}, {Adams}, \& {Bergin}}]{Cleeves13a}
{Cleeves}, L.~I., {Adams}, F.~C., \& {Bergin}, E.~A. 2013, \apj, 772, 5

\bibitem[{{Cleeves} {et~al.}(2015){Cleeves}, {Bergin}, {Qi}, {Adams}, \&
  {{\"O}berg}}]{Cleeves15}
{Cleeves}, L.~I., {Bergin}, E.~A., {Qi}, C., {Adams}, F.~C., \& {{\"O}berg},
  K.~I. 2015, \apj, 799, 204

\bibitem[{{D'Alessio} {et~al.}(1999){D'Alessio}, {Calvet}, {Hartmann},
  {Lizano}, \& {Cant{\'o}}}]{dalessio99}
{D'Alessio}, P., {Calvet}, N., {Hartmann}, L., {Lizano}, S., \& {Cant{\'o}}, J.
  1999, \apj, 527, 893

\bibitem[{{Dartois} {et~al.}(2003){Dartois}, {Dutrey}, \& {Guilloteau}}]{ddg03}
{Dartois}, E., {Dutrey}, A., \& {Guilloteau}, S. 2003, \aap, 399, 773

\bibitem[{{Debes} {et~al.}(2013){Debes}, {Jang-Condell}, {Weinberger},
  {Roberge}, \& {Schneider}}]{Debes13}
{Debes}, J.~H., {Jang-Condell}, H., {Weinberger}, A.~J., {Roberge}, A., \&
  {Schneider}, G. 2013, \apj, 771, 45

\bibitem[{{Draine}(2006)}]{Draine06}
{Draine}, B.~T. 2006, \apj, 636, 1114

\bibitem[{{Draine} \& {Lee}(1984)}]{dl84}
{Draine}, B.~T., \& {Lee}, H.~M. 1984, \apj, 285, 89

\bibitem[{{Du} \& {Bergin}(2014)}]{Du14}
{Du}, F., \& {Bergin}, E.~A. 2014, \apj, 792, 2

\bibitem[{{Du} {et~al.}(2015){Du}, {Bergin}, \& {Hogerheijde}}]{Du15}
{Du}, F., {Bergin}, E.~A., \& {Hogerheijde}, M.~R. 2015, \apjl, 807, L32

\bibitem[{{Dullemond} \& {Dominik}(2004)}]{dd04}
{Dullemond}, C.~P., \& {Dominik}, C. 2004, \aap, 421, 1075

\bibitem[{{Dullemond} {et~al.}(2002){Dullemond}, {van Zadelhoff}, \&
  {Natta}}]{Dullemond2002}
{Dullemond}, C.~P., {van Zadelhoff}, G.~J., \& {Natta}, A. 2002, \aap, 389, 464

\bibitem[{{Dutrey} {et~al.}(1997){Dutrey}, {Guilloteau}, \& {Guelin}}]{dgg97}
{Dutrey}, A., {Guilloteau}, S., \& {Guelin}, M. 1997, \aap, 317, L55

\bibitem[{{Dutrey} {et~al.}(2003){Dutrey}, {Guilloteau}, \& {Simon}}]{Dutrey03}
{Dutrey}, A., {Guilloteau}, S., \& {Simon}, M. 2003, \aap, 402, 1003

\bibitem[{{Espaillat} {et~al.}(2010){Espaillat}, {D'Alessio}, {Hern{\'a}ndez},
  {Nagel}, {Luhman}, {Watson}, {Calvet}, {Muzerolle}, \&
  {McClure}}]{espaillat2010}
{Espaillat}, C., {D'Alessio}, P., {Hern{\'a}ndez}, J., {et~al.} 2010, \apj,
  717, 441

\bibitem[{{Favre} {et~al.}(2013){Favre}, {Cleeves}, {Bergin}, {Qi}, \&
  {Blake}}]{favre13a}
{Favre}, C., {Cleeves}, L.~I., {Bergin}, E.~A., {Qi}, C., \& {Blake}, G.~A.
  2013, \apjl, 776, L38

\bibitem[{{Fogel} {et~al.}(2011){Fogel}, {Bethell}, {Bergin}, {Calvet}, \&
  {Semenov}}]{Fogel11}
{Fogel}, J.~K.~J., {Bethell}, T.~J., {Bergin}, E.~A., {Calvet}, N., \&
  {Semenov}, D. 2011, \apj, 726, 29

\bibitem[{{Furlan} {et~al.}(2006)}]{furlan06}
{Furlan}, E., {et~al.} 2006, \apjs, 165, 568

\bibitem[{{Furuya} \& {Aikawa}(2014)}]{Furuya14}
{Furuya}, K., \& {Aikawa}, Y. 2014, \apj, 790, 97

\bibitem[{{Geers} {et~al.}(2007){Geers}, {van Dishoeck}, {Visser},
  {Pontoppidan}, {Augereau}, {Habart}, \& {Lagrange}}]{geers07}
{Geers}, V.~C., {van Dishoeck}, E.~F., {Visser}, R., {et~al.} 2007, \aap, 476,
  279

\bibitem[{{Goldsmith} {et~al.}(2008){Goldsmith}, {Heyer}, {Narayanan}, {Snell},
  {Li}, \& {Brunt}}]{pg_taurus}
{Goldsmith}, P.~F., {Heyer}, M., {Narayanan}, G., {et~al.} 2008, \apj, 680, 428

\bibitem[{{Guilloteau} \& {Dutrey}(1998)}]{gd98}
{Guilloteau}, S., \& {Dutrey}, A. 1998, \aap, 339, 467

\bibitem[{{Gundlach} \& {Blum}(2015)}]{Gundlach15}
{Gundlach}, B., \& {Blum}, J. 2015, \apj, 798, 34

\bibitem[{{Habing}(1968)}]{habing68}
{Habing}, H.~J. 1968, \bain, 19, 421

\bibitem[{{Hartmann} {et~al.}(2001){Hartmann}, {Ballesteros-Paredes}, \&
  {Bergin}}]{hbb01}
{Hartmann}, L., {Ballesteros-Paredes}, J., \& {Bergin}, E.~A. 2001, \apj, 562,
  852

\bibitem[{{Hartmann} {et~al.}(1998{\natexlab{a}}){Hartmann}, {Calvet},
  {Gullbring}, \& {D'Alessio}}]{Hartmann98}
{Hartmann}, L., {Calvet}, N., {Gullbring}, E., \& {D'Alessio}, P.
  1998{\natexlab{a}}, \apj, 495, 385

\bibitem[{{Hartmann} {et~al.}(1998{\natexlab{b}}){Hartmann}, {Calvet},
  {Gullbring}, \& {D'Alessio}}]{Hartmann1998}
---. 1998{\natexlab{b}}, \apj, 495, 385

\bibitem[{{Henning} {et~al.}(2010){Henning}, {Semenov}, {Guilloteau}, {Dutrey},
  {Hersant}, {Wakelam}, {Chapillon}, {Launhardt}, {Pi{\'e}tu}, \&
  {Schreyer}}]{Henning10}
{Henning}, T., {Semenov}, D., {Guilloteau}, S., {et~al.} 2010, \apj, 714, 1511

\bibitem[{{Herczeg} {et~al.}(2002){Herczeg}, {Linsky}, {Valenti},
  {Johns-Krull}, \& {Wood}}]{herczeg_twhya1}
{Herczeg}, G.~J., {Linsky}, J.~L., {Valenti}, J.~A., {Johns-Krull}, C.~M., \&
  {Wood}, B.~E. 2002, \apj, 572, 310

\bibitem[{{Herczeg} {et~al.}(2004){Herczeg}, {Wood}, {Linsky}, {Valenti}, \&
  {Johns-Krull}}]{herczeg_twhya2}
{Herczeg}, G.~J., {Wood}, B.~E., {Linsky}, J.~L., {Valenti}, J.~A., \&
  {Johns-Krull}, C.~M. 2004, \apj, 607, 369

\bibitem[{{Hogerheijde} {et~al.}(2016){Hogerheijde}, {Bekkers}, {Pinilla},
  {Salinas}, {Kama}, {Andrews}, {Qi}, \& {Wilner}}]{Hogerheijde16}
{Hogerheijde}, M.~R., {Bekkers}, D., {Pinilla}, P., {et~al.} 2016, \aap, 586,
  A99

\bibitem[{{Hogerheijde} {et~al.}(2015){Hogerheijde}, {Bekkers}, {Pinilla},
  {Vachail}, {Kama}, {Andrews}, {Qi}, \& {Wilner}}]{Hogerheijde15}
---. 2015, \aap, submitted

\bibitem[{{Hogerheijde} {et~al.}(2011){Hogerheijde}, {Bergin}, {Brinch},
  {et~al.}}]{hoger11a}
{Hogerheijde}, M.~R., {Bergin}, E.~A., {Brinch}, C., {et~al.} 2011, Science,
  334, 338

\bibitem[{{Hollenbach} {et~al.}(2009){Hollenbach}, {Kaufman}, {Bergin}, \&
  {Melnick}}]{hkbm09}
{Hollenbach}, D., {Kaufman}, M.~J., {Bergin}, E.~A., \& {Melnick}, G.~J. 2009,
  \apj, 690, 1497

\bibitem[{{Hueso} \& {Guillot}(2005)}]{Hueso05}
{Hueso}, R., \& {Guillot}, T. 2005, \aap, 442, 703

\bibitem[{{Isella} {et~al.}(2007){Isella}, {Testi}, {Natta}, {Neri}, {Wilner},
  \& {Qi}}]{isella07}
{Isella}, A., {Testi}, L., {Natta}, A., {et~al.} 2007, \aap, 469, 213

\bibitem[{{Johansen} {et~al.}(2007){Johansen}, {Oishi}, {Mac Low}, {Klahr},
  {Henning}, \& {Youdin}}]{Johansen07}
{Johansen}, A., {Oishi}, J.~S., {Mac Low}, M.-M., {et~al.} 2007, \nat, 448,
  1022

\bibitem[{{Jones} {et~al.}(2013){Jones}, {Fanciullo}, {K{\"o}hler},
  {Verstraete}, {Guillet}, {Bocchio}, \& {Ysard}}]{Jones13}
{Jones}, A.~P., {Fanciullo}, L., {K{\"o}hler}, M., {et~al.} 2013, \aap, 558,
  A62

\bibitem[{{Kama} {et~al.}(2016{\natexlab{a}}){Kama}, {Bruderer}, {Carney},
  {Hogerheijde}, {van Dishoeck}, {Fedele}, {Baryshev}, {Boland}, {G{\"u}sten},
  {Aikutalp}, {Choi}, {Endo}, {Frieswijk}, {Karska}, {Klaassen}, {Koumpia},
  {Kristensen}, {Leurini}, {Nagy}, {Perez Beaupuits}, {Risacher}, {van der
  Marel}, {van Kempen}, {van Weeren}, {Wyrowski}, \& {Y{\i}ld{\i}z}}]{Kama16b}
{Kama}, M., {Bruderer}, S., {Carney}, M., {et~al.} 2016{\natexlab{a}}, \aap, in
  press

\bibitem[{{Kama} {et~al.}(2016{\natexlab{b}}){Kama}, {Bruderer}, {van
  Dishoeck}, {Hogerheijde}, {Folsom}, {Miotello}, {Fedele}, {Belloche},
  {G{\"u}sten}, \& {Wyrowski}}]{Kama16a}
{Kama}, M., {Bruderer}, S., {van Dishoeck}, E.~F., {et~al.} 2016{\natexlab{b}},
  \aap, in press.

\bibitem[{{Kastner} {et~al.}(2014){Kastner}, {Hily-Blant}, {Rodriguez},
  {Punzi}, \& {Forveille}}]{Kastner14}
{Kastner}, J.~H., {Hily-Blant}, P., {Rodriguez}, D.~R., {Punzi}, K., \&
  {Forveille}, T. 2014, \apj, 793, 55

\bibitem[{{Kastner} {et~al.}(2015){Kastner}, {Qi}, {Gorti}, {Hily-Blant},
  {Oberg}, {Forveille}, {Andrews}, \& {Wilner}}]{Kastner15}
{Kastner}, J.~H., {Qi}, C., {Gorti}, U., {et~al.} 2015, \apj, 806, 75

\bibitem[{{Kenyon} {et~al.}(1994){Kenyon}, {Dobrzycka}, \&
  {Hartmann}}]{Kenyon94}
{Kenyon}, S.~J., {Dobrzycka}, D., \& {Hartmann}, L. 1994, \aj, 108, 1872

\bibitem[{{Kitamura} {et~al.}(2002){Kitamura}, {Momose}, {Yokogawa}, {Kawabe},
  {Tamura}, \& {Ida}}]{Kitamura02}
{Kitamura}, Y., {Momose}, M., {Yokogawa}, S., {et~al.} 2002, \apj, 581, 357

\bibitem[{{Lynden-Bell} \& {Pringle}(1974)}]{LyndenBell1974}
{Lynden-Bell}, D., \& {Pringle}, J.~E. 1974, \mnras, 168, 603

\bibitem[{{Mart{\'{\i}}n-Dom{\'e}nech}
  {et~al.}(2014){Mart{\'{\i}}n-Dom{\'e}nech}, {Mu{\~n}oz Caro}, {Bueno}, \&
  {Goesmann}}]{Martin-Domenench14}
{Mart{\'{\i}}n-Dom{\'e}nech}, R., {Mu{\~n}oz Caro}, G.~M., {Bueno}, J., \&
  {Goesmann}, F. 2014, \aap, 564, A8

\bibitem[{{Mathis} {et~al.}(1977){Mathis}, {Rumpl}, \& {Nordsieck}}]{mrn}
{Mathis}, J.~S., {Rumpl}, W., \& {Nordsieck}, K.~H. 1977, \apj, 217, 425

\bibitem[{{McClure} {et~al.}(2010){McClure}, {Furlan}, {Manoj}, {Luhman},
  {Watson}, {Forrest}, {Espaillat}, {Calvet}, {D'Alessio}, {Sargent}, {Tobin},
  \& {Chiang}}]{McClure10}
{McClure}, M.~K., {Furlan}, E., {Manoj}, P., {et~al.} 2010, \apjs, 188, 75

\bibitem[{{McClure} {et~al.}(2016){McClure}, {Bergin}, {Cleeves}, {van
  Dishoeck}, {Blake}, {Evans}, {Green}, {Henning}, {\"{O}berg}, {Pontoppidan},
  \& {Salyk}}]{McClure16}
{McClure}, M.~K., {Bergin}, E.~A., {Cleeves}, L.~I., {et~al.} 2016, \apj,
  submitted

\bibitem[{{Menu} {et~al.}(2014){Menu}, {van Boekel}, {Henning}, {Chandler},
  {Linz}, {Benisty}, {Lacour}, {Min}, {Waelkens}, {Andrews}, {Calvet},
  {Carpenter}, {Corder}, {Deller}, {Greaves}, {Harris}, {Isella}, {Kwon},
  {Lazio}, {Le Bouquin}, {M{\'e}nard}, {Mundy}, {P{\'e}rez}, {Ricci},
  {Sargent}, {Storm}, {Testi}, \& {Wilner}}]{Menu14}
{Menu}, J., {van Boekel}, R., {Henning}, T., {et~al.} 2014, \aap, 564, A93

\bibitem[{{Mumma} \& {Charnley}(2011)}]{mc11}
{Mumma}, M.~J., \& {Charnley}, S.~B. 2011, \araa, 49, 471

\bibitem[{{Nagy} {et~al.}(2015){Nagy}, {Ossenkopf}, {Van der Tak}, {Faure},
  {Makai}, \& {Bergin}}]{Nagy15}
{Nagy}, Z., {Ossenkopf}, V., {Van der Tak}, F.~F.~S., {et~al.} 2015, \aap, 578,
  A124

\bibitem[{{Nomura} {et~al.}(2016){Nomura}, {Tsukagoshi}, {Kawabe}, {Ishimoto},
  {Okuzumi}, {Muto}, {Kanagawa}, {Ida}, {Walsh}, {Millar}, \& {Bai}}]{Nomura16}
{Nomura}, H., {Tsukagoshi}, T., {Kawabe}, R., {et~al.} 2016, \apjl, 819, L7

\bibitem[{{{\"O}berg} {et~al.}(2011{\natexlab{a}}){{\"O}berg}, {Boogert},
  {Pontoppidan}, {van den Broek}, {van Dishoeck}, {Bottinelli}, {Blake}, \&
  {Evans}}]{Oberg11c}
{{\"O}berg}, K.~I., {Boogert}, A.~C.~A., {Pontoppidan}, K.~M., {et~al.}
  2011{\natexlab{a}}, ApJ, 740, 109

\bibitem[{{{\"O}berg} {et~al.}(2011{\natexlab{b}}){{\"O}berg}, {Murray-Clay},
  \& {Bergin}}]{omb11}
{{\"O}berg}, K.~I., {Murray-Clay}, R., \& {Bergin}, E.~A. 2011{\natexlab{b}},
  \apjl, 743, L16

\bibitem[{{Pani{\'c}} {et~al.}(2009){Pani{\'c}}, {Hogerheijde}, {Wilner}, \&
  {Qi}}]{panic09}
{Pani{\'c}}, O., {Hogerheijde}, M.~R., {Wilner}, D., \& {Qi}, C. 2009, \aap,
  501, 269

\bibitem[{{P{\'e}rez} {et~al.}(2015){P{\'e}rez}, {Chandler}, {Isella},
  {Carpenter}, {Andrews}, {Calvet}, {Corder}, {Deller}, {Dullemond}, {Greaves},
  {Harris}, {Henning}, {Kwon}, {Lazio}, {Linz}, {Mundy}, {Ricci}, {Sargent},
  {Storm}, {Tazzari}, {Testi}, \& {Wilner}}]{Perez15}
{P{\'e}rez}, L.~M., {Chandler}, C.~J., {Isella}, A., {et~al.} 2015, \apj, 813,
  41

\bibitem[{{Pi{\'e}tu} {et~al.}(2007){Pi{\'e}tu}, {Dutrey}, \&
  {Guilloteau}}]{pdg07}
{Pi{\'e}tu}, V., {Dutrey}, A., \& {Guilloteau}, S. 2007, \aap, 467, 163

\bibitem[{{Pilleri} {et~al.}(2013){Pilleri}, {Trevi{\~n}o-Morales}, {Fuente},
  {Joblin}, {Cernicharo}, {Gerin}, {Viti}, {Bern{\'e}}, {Goicoechea}, {Pety},
  {Gonzalez-Garc{\'{\i}}a}, {Montillaud}, {Ossenkopf}, {Kramer},
  {Garc{\'{\i}}a-Burillo}, {Le Petit}, \& {Le Bourlot}}]{Pilleri13}
{Pilleri}, P., {Trevi{\~n}o-Morales}, S., {Fuente}, A., {et~al.} 2013, \aap,
  554, A87

\bibitem[{{Piso} {et~al.}(2015){Piso}, {{\"O}berg}, {Birnstiel}, \&
  {Murray-Clay}}]{piso15}
{Piso}, A.-M.~A., {{\"O}berg}, K.~I., {Birnstiel}, T., \& {Murray-Clay}, R.~A.
  2015, \apj, 815, 109

\bibitem[{{Qi} {et~al.}(2013){Qi}, {{\"O}berg}, {Wilner}, {D'Alessio},
  {Bergin}, {Andrews}, {Blake}, {Hogerheijde}, \& {van Dishoeck}}]{Qi13a}
{Qi}, C., {{\"O}berg}, K.~I., {Wilner}, D.~J., {et~al.} 2013, Science, 341, 630

\bibitem[{{Raassen}(2009)}]{Raassen09}
{Raassen}, A.~J.~J. 2009, \aap, 505, 755

\bibitem[{{Reboussin} {et~al.}(2015){Reboussin}, {Wakelam}, {Guilloteau},
  {Hersant}, \& {Dutrey}}]{Reboussin15}
{Reboussin}, L., {Wakelam}, V., {Guilloteau}, S., {Hersant}, F., \& {Dutrey},
  A. 2015, \aap, 579, A82

\bibitem[{{Schindhelm} {et~al.}(2012){Schindhelm}, {France}, {Herczeg},
  {Bergin}, {Yang}, {Brown}, {Brown}, {Linsky}, \& {Valenti}}]{Schindhelm12}
{Schindhelm}, E., {France}, K., {Herczeg}, G.~J., {et~al.} 2012, \apjl, 756,
  L23

\bibitem[{{Schlafly} {et~al.}(2014){Schlafly}, {Green}, {Finkbeiner}, {Rix},
  {Bell}, {Burgett}, {Chambers}, {Draper}, {Hodapp}, {Kaiser}, {Magnier},
  {Martin}, {Metcalfe}, {Price}, \& {Tonry}}]{Schlafly14}
{Schlafly}, E.~F., {Green}, G., {Finkbeiner}, D.~P., {et~al.} 2014, \apj, 786,
  29

\bibitem[{{Schneider} {et~al.}(2005){Schneider}, {Silverstone}, {Hines},
  {Cotera}, {Grady}, {Stapelfeldt}, {Padgett}, {Menard}, {Wolf}, \&
  {Stecklum}}]{Schneider05}
{Schneider}, G., {Silverstone}, M.~D., {Hines}, M.~D., {et~al.} 2005, in
  Protostars and Planets V Posters, 8540

\bibitem[{{Schwarz} {et~al.}(2016){Schwarz}, {Bergin}, {Cleeves}, {Blake},
  {Zhang}, {{\"O}berg}, {van Dishoeck}, \& {Qi}}]{Schwarz16}
{Schwarz}, K.~R., {Bergin}, E.~A., {Cleeves}, L.~I., {et~al.} 2016, \apj, 823,
  91

\bibitem[{{Semenov} \& {Wiebe}(2011)}]{Semenov11}
{Semenov}, D., \& {Wiebe}, D. 2011, \apjs, 196, 25

\bibitem[{{Simon} {et~al.}(2000){Simon}, {Dutrey}, \& {Guilloteau}}]{sdg00}
{Simon}, M., {Dutrey}, A., \& {Guilloteau}, S. 2000, \apj, 545, 1034

\bibitem[{{Spielfiedel} {et~al.}(2012){Spielfiedel}, {Feautrier}, {Najar}, {Ben
  Abdallah}, {Dayou}, {Senent}, \& {Lique}}]{c2hcolxs}
{Spielfiedel}, A., {Feautrier}, N., {Najar}, F., {et~al.} 2012, \mnras, 421,
  1891

\bibitem[{{Tachihara} {et~al.}(2009){Tachihara}, {Neuh{\"a}user}, \&
  {Fukui}}]{tachihara09}
{Tachihara}, K., {Neuh{\"a}user}, R., \& {Fukui}, Y. 2009, \pasj, 61, 585

\bibitem[{{Thi} {et~al.}(2010){Thi}, {Mathews}, {M{\'e}nard}, {Woitke},
  {Meeus}, {Riviere-Marichalar}, {Pinte}, {Howard}, {Roberge}, {Sandell},
  {Pascucci}, {Riaz}, {Grady}, {Dent}, {Kamp}, {Duch{\^e}ne}, {Augereau},
  {Pantin}, {Vandenbussche}, {Tilling}, {Williams}, {Eiroa}, {Barrado},
  {Alacid}, {Andrews}, {Ardila}, {Aresu}, {Brittain}, {Ciardi}, {Danchi},
  {Fedele}, {de Gregorio-Monsalvo}, {Heras}, {Huelamo}, {Krivov}, {Lebreton},
  {Liseau}, {Martin-Zaidi}, {Mendigut{\'{\i}}a}, {Montesinos}, {Mora},
  {Morales-Calderon}, {Nomura}, {Phillips}, {Podio}, {Poelman}, {Ramsay},
  {Rice}, {Solano}, {Walker}, {White}, \& {Wright}}]{thi10}
{Thi}, W.-F., {Mathews}, G., {M{\'e}nard}, F., {et~al.} 2010, \aap, 518, 647

\bibitem[{{Tielens}(2005)}]{tielens_book}
{Tielens}, A.~G.~G.~M. 2005, {The Physics and Chemistry of the Interstellar
  Medium} (Cambridge University Press, 2005.)

\bibitem[{{Tielens} \& {Hollenbach}(1985)}]{th85}
{Tielens}, A.~G.~G.~M., \& {Hollenbach}, D. 1985, \apj, 291, 722

\bibitem[{{Tsukagoshi} {et~al.}(2015){Tsukagoshi}, {Momose}, {Saito},
  {Kitamura}, {Shimajiri}, \& {Kawabe}}]{Tsukagoshi15}
{Tsukagoshi}, T., {Momose}, M., {Saito}, M., {et~al.} 2015, \apjl, 802, L7

\bibitem[{{Tsukagoshi} {et~al.}(2016){Tsukagoshi}, {Nomura}, {Muto}, {Kawabe},
  {Ishimoto}, {Kanagawa}, {Okuzumi}, {Ida}, {Walsh}, \&
  {Millar}}]{Tsukagoshi16}
{Tsukagoshi}, T., {Nomura}, H., {Muto}, T., {et~al.} 2016, \apjl, in press.

\bibitem[{{Vacca} \& {Sandell}(2011)}]{vs11}
{Vacca}, W.~D., \& {Sandell}, G. 2011, \apj, 732, 8

\bibitem[{{van der Tak} {et~al.}(2007){van der Tak}, {Black}, {Sch{\"o}ier},
  {Jansen}, \& {van Dishoeck}}]{radex}
{van der Tak}, F.~F.~S., {Black}, J.~H., {Sch{\"o}ier}, F.~L., {Jansen}, D.~J.,
  \& {van Dishoeck}, E.~F. 2007, \aap, 468, 627

\bibitem[{{van Leeuwen}(2007)}]{vanLeeuwen07}
{van Leeuwen}, F. 2007, \aap, 474, 653

\bibitem[{{Visser} {et~al.}(2007){Visser}, {Geers}, {Dullemond}, {Augereau},
  {Pontoppidan}, \& {van Dishoeck}}]{Visser07}
{Visser}, R., {Geers}, V.~C., {Dullemond}, C.~P., {et~al.} 2007, \aap, 466, 229

\bibitem[{{Weidenschilling} \& {Cuzzi}(1993)}]{wc_ppiii}
{Weidenschilling}, S.~J., \& {Cuzzi}, J.~N. 1993, in Protostars and Planets
  III, ed. E.~H. {Levy} \& J.~I. {Lunine}, 1031--1060

\bibitem[{{Weinberger} {et~al.}(2002){Weinberger}, {Becklin}, {Schneider},
  {Chiang}, {Lowrance}, {Silverstone}, {Zuckerman}, {Hines}, \&
  {Smith}}]{weinberger02}
{Weinberger}, A.~J., {Becklin}, E.~E., {Schneider}, G., {et~al.} 2002, \apj,
  566, 409

\bibitem[{{Whipple}(1972)}]{whipple1972}
{Whipple}, F.~L. 1972, in From Plasma to Planet, ed. A.~{Elvius}, 211

\bibitem[{{Whittet}(2010)}]{Whittet10}
{Whittet}, D.~C.~B. 2010, \apj, 710, 1009

\bibitem[{{Williams} \& {Best}(2014)}]{jWilliams14}
{Williams}, J.~P., \& {Best}, W.~M.~J. 2014, \apj, 788, 59

\bibitem[{{Yang} {et~al.}(2012){Yang}, {Herczeg}, {Linsky}, {Brown},
  {Johns-Krull}, {Ingleby}, {Calvet}, {Bergin}, \& {Valenti}}]{yang12}
{Yang}, H., {Herczeg}, G.~J., {Linsky}, J.~L., {et~al.} 2012, \apj, 744, 121

\bibitem[{{Youdin} \& {Kenyon}(2013)}]{Youdin13}
{Youdin}, A.~N., \& {Kenyon}, S.~J. 2013, {From Disks to Planets}, ed. T.~D.
  {Oswalt}, L.~M. {French}, \& P.~{Kalas}, 1

\bibitem[{{Zhang} {et~al.}(2016){Zhang}, {Bergin}, {Blake}, {Cleeves},
  {Hogerheijde}, {Salinas}, \& {Schwarz}}]{Zhang16}
{Zhang}, K., {Bergin}, E.~A., {Blake}, G.~A., {et~al.} 2016, \apjl, 818, L16

\bibitem[{{Zhang} {et~al.}(2013){Zhang}, {Pontoppidan}, {Salyk}, \&
  {Blake}}]{Zhang13}
{Zhang}, K., {Pontoppidan}, K.~M., {Salyk}, C., \& {Blake}, G.~A. 2013, \apj,
  766, 82

\end{thebibliography}

\end{document}